\documentclass[twocolumn]{emulateapj}

\usepackage{bm}

\usepackage{subfigure}

\usepackage{amsmath, amssymb}
\usepackage{times}

\usepackage[usenames]{xcolor}
\definecolor{mpl_red}{HTML}{D62728}

\usepackage[backref, breaklinks, plainpages=false, colorlinks=true, anchorcolor=blue!50!black, citecolor=blue!50!black, linkcolor=blue!50!black, urlcolor=mpl_red, bookmarks=false]{hyperref}
\usepackage{natbib}

\journalinfo{\textsc{Accepted for publication in The Astrophysical Journal}\textup{, 2018~August~30}}
\submitted{Accepted for publication in The Astrophysical Journal on 2018~August~30}

\shorttitle{PULSE MORPHOLOGY OF PSR~J1745--2900}
\shortauthors{PEARLMAN ET AL.}

\begin{document}

\renewcommand*{\backref}[1]{[#1]}

\title{Pulse Morphology of the Galactic Center Magnetar PSR~J1745--2900}

\newcommand{\Pearlman}{Aaron~B.~Pearlman}
\newcommand{\Majid}{Walid~A.~Majid}
\newcommand{\Prince}{Thomas~A.~Prince}
\newcommand{\Kocz}{Jonathon~Kocz}
\newcommand{\Horiuchi}{Shinji~Horiuchi}

\newcommand{\CaltechPhysics}{Division of Physics, Mathematics, and Astronomy, California Institute of Technology, Pasadena, CA 91125, USA; \textcolor{blue}{aaron.b.pearlman@caltech.edu}}
\newcommand{\JPL}{Jet Propulsion Laboratory, California Institute of Technology, Pasadena, CA 91109, USA}
\newcommand{\CDSCC}{CSIRO Astronomy and Space Science, Canberra Deep Space Communications Complex, P.O.~Box~1035, Tuggeranong, ACT~2901, Australia}
\newcommand{\NDSEG}{$^{\text{4}}$~NDSEG Research Fellow.}
\newcommand{\NSF}{$^{\text{5}}$~NSF Graduate Research Fellow.}

\author{\Pearlman\altaffilmark{1,4,5}, \Majid\altaffilmark{2,1}, \Prince\altaffilmark{1,2}, \Kocz\altaffilmark{1}, and~\Horiuchi\altaffilmark{3}}

\address{
	$^{\text{1}}$~\CaltechPhysics \\
	$^{\text{2}}$~\JPL \\
	$^{\text{3}}$~\CDSCC}

\thanks{\NDSEG}
\thanks{\NSF}

\begin{abstract}

We present results from observations of the Galactic Center magnetar, \text{PSR~J1745--2900}, at 2.3~and~8.4\,GHz with the NASA Deep Space Network 70\,m~antenna,~DSS-43. We study the magnetar's radio profile shape, flux density, radio spectrum, and single pulse behavior over a $\sim$1~year period between MJDs~57233~and~57621. In particular, the magnetar exhibits a significantly negative average spectral index of \mbox{$\langle\alpha\rangle$\,$=$\,--1.86\,$\pm$\,0.02} when the 8.4\,GHz profile is \text{single-peaked}, which flattens considerably when the profile is \text{double-peaked}. We have carried out an analysis of single pulses at 8.4\,GHz on MJD~57479 and find that giant pulses and pulses with multiple emission components are emitted during a significant number of rotations. The resulting single pulse flux density distribution is incompatible with a \text{log-normal} distribution. The typical pulse width of the components is $\sim$1.8\,ms, and the prevailing delay time between successive components is $\sim$7.7\,ms. Many of the single pulse emission components show significant frequency structure over bandwidths of $\sim$100\,MHz, which we believe is the first observation of such behavior from a radio magnetar. We report a characteristic single pulse broadening timescale of $\langle\tau_{d}\rangle$\,$=$\,6.9\,$\pm$\,0.2\,ms at 8.4\,GHz. We find that the pulse broadening is highly variable between emission components and cannot be explained by a thin scattering screen at distances~$\gtrsim$\,1\,kpc. We discuss possible intrinsic and extrinsic mechanisms for the magnetar's emission and compare our results to other magnetars, high magnetic field pulsars, and fast radio bursts.

\end{abstract}

\keywords{Galaxy: center -- pulsars: individual (PSR~J1745--2900) -- scattering -- stars: magnetars -- stars: neutron}


\section{Introduction}

\setcounter{footnote}{5}

Magnetars are a class of slowly rotating neutron stars, with spin periods between $\sim$2 and 12\,s, that are thought to be powered by their decaying ultra-strong magnetic fields~\citep{Duncan1992, Thompson1995, Thompson1996}. More than $\sim$2600~pulsars have now been found, but only 31~magnetars or magnetar candidates are currently known~(\citealt{Kaspi2017};~see~the~McGill Magnetar Catalog\footnote{http://www.physics.mcgill.ca/$\sim$pulsar/magnetar/main.html.}). Most of these are Galactic magnetars, many of which are located in the inner region of the Milky~Way~\citep{Olausen2014}. Typical surface dipolar magnetic fields of magnetars range between $\sim$10$^{\text{14}}$ and 10$^{\text{15}}$\,G, which exceed the $\sim$10$^{\text{12}}$\,G fields of \text{rotation-powered} pulsars. Transient \text{X-ray} and \text{gamma-ray} outbursts are hallmark features of magnetar emission and have led to the discovery of the majority of new magnetars.

\text{PSR~J1745--2900} is one of only four magnetars with detectable radio pulsations~\citep{Camilo2006, Camilo2007a, Levin2010, Eatough2013b, Shannon2013}. It is unique among the population of magnetars because of its close proximity to the 4\,$\times$\,10$^{\text{6}}$\,$M_{\odot}$ black hole, Sagittarius A$^{*}$~(Sgr~A$^{*}$), at the Galactic Center~(GC). The discovery of a rare magnetar near the~GC may suggest that the environment around Sgr~A$^{*}$ is more conducive for magnetar formation~\citep{Dexter2014}. Observations of \text{PSR~J1745--2900} also provide a valuable probe of the interstellar medium~(ISM) near the~GC~(e.g.,~\citealt{Eatough2013b, Bower2014, Spitler2014, Desvignes2018}), which may shed light on why previous searches for radio pulsars within $\sim$10\,arcmin of Sgr~A$^{*}$ have been unsuccessful~\citep{Kramer2000, Johnston2006, Deneva2009, Macquart2010, Bates2011, Eatough2013a, Siemion2013}. It is widely believed that these searches may have been hindered by \text{scattering-induced} pulse broadening of the pulsed radio emission as a result of large electron densities along the line of sight.

The GC~magnetar was serendipitiously discovered by the \textit{Swift}\footnote{The \textit{Swift} \text{Gamma-Ray} Burst Explorer was renamed the ``Neil~Gehrels \textit{Swift} Observatory'' in honor of Neil~Gehrels, \textit{Swift's} principal investigator.} Burst Alert Telescope~(BAT) following an \text{X-ray} flare near Sgr~A$^{*}$ and is the most recent addition to the radio magnetar family~\citep{Eatough2013b, Kennea2013, Shannon2013}. Subsequent observations with the \textit{NuSTAR} \text{X-ray} telescope uncovered \text{X-ray} pulsations at a period of $P$\,$=$\,3.76\,s and a \text{spin-down} rate of $\dot{P}$\,$=$\,6.5\,$\times$\,10$^{\text{--12}}$\,s\,s$^{\text{--1}}$~\citep{Mori2013}. Assuming a dipolar magnetic field, this implies a surface magnetic field of $B_{\text{surf}}$\,$\approx$\,1.6\,$\times$\,10$^{\text{14}}$\,G, \text{spin-down} luminosity of $\dot{E}$\,$\approx$\,5\,$\times$\,10$^{\text{33}}$\,erg\,s$^{\text{--1}}$, and characteristic age of $\tau_{\text{c}}$\,$\approx$\,9\,kyr. A series of \textit{Chandra} and \textit{Swift} observations were later performed, which localized the magnetar to an angular distance of 2.4\,arcsec from Sgr~A$^{*}$~\citep{Rea2013}. The proper motion of \text{PSR~J1745--2900} was measured relative to Sgr~A$^{*}$ using the Very Long Baseline Array~(VLBA), which yielded a transverse velocity of 236\,km\,s$^{\text{--1}}$ at a projected separation of 0.097\,pc~\citep{Bower2015}.

Radio pulsations have been detected from \text{PSR~J1745--2900} at frequencies between 1.2~and~291\,GHz, and its radio spectrum is relatively flat~\citep{Eatough2013b, Spitler2014, Torne2015, Torne2017}. Multifrequency radio observations established that the GC~magnetar has the largest dispersion measure~(DM\,$=$\,1778\,$\pm$\,3\,pc\,cm$^{\text{--3}}$) and Faraday rotation measure~(\mbox{RM\,$=$\,--66,960\,$\pm$\,50\,rad\,m$^{\text{--2}}$}) of any known pulsar~\citep{Eatough2013b}. \citet{Schnitzeler2016} found that its~RM had increased to \mbox{--66,080\,$\pm$\,24\,rad\,m$^{\text{--2}}$} approximately 2~years later, and recent measurements by~\citet{Desvignes2018} showed that its linear polarization fraction and~RM were both significantly variable over a time span of roughly 4~years.

Single pulse radio observations of \text{PSR~1745--2900} have been performed at 8.7\,GHz by~\citet{Lynch2015} with the Green Bank Telescope~(GBT) and at 8.6\,GHz by~\citet{Yan2015} using the Shanghai Tian Ma Radio Telescope~(TMRT). \citet{Lynch2015} showed that the magnetar experienced a transition from a stable state to a more erratic state early in 2014. During this period, significant changes in the magnetar's flux density, radio profile shape, and single pulse properties were observed. \citet{Yan2015} presented single pulse observations between 2014~June~28 and October~13, and they performed an analysis of pulses detected during an erratic period on MJD~56836. \citet{Yan2018} recently reported on single pulse observations at 3.1\,GHz with the Parkes radio telescope, which showed that the magnetar was in a stable state between MJDs~56475~and~56514.

Temporal scatter broadening measurements were performed by~\citet{Spitler2014} using single pulses and average pulse profiles from \text{PSR~J1745--2900} between 1.19~and~18.95\,GHz. They derived a pulse broadening spectral index of $\alpha_{d}$\,$=$\,--3.8\,$\pm$\,0.2 and a pulse broadening timescale of $\tau_{d}$\,$=$\,1.3\,$\pm$\,0.2\,s at 1\,GHz, which is several orders of magnitude lower than the value predicted by the NE2001 electron density model~\citep{Cordes2002}. Observations with the~VLBA and phased array of the Karl~G.~Jansky Very Large Array~(VLA) were subsequently performed to measure the angular broadening of \text{PSR~J1745--2900}~\citep{Bower2014}. \citet{Bower2014} argued that the observed scattering is consistent with a single thin screen at a distance of $\Delta_{\text{GC}}$\,$=$\,5.8\,$\pm$\,0.3\,kpc from the~GC. A secondary scattering screen, located $\sim$0.1\,pc in front of the magnetar, was recently proposed by~\citet{Desvignes2018} to explain the magnetar's depolarization at low radio frequencies.

In this paper, we present results from simultaneous observations of \text{PSR~J1745--2900} at 2.3~and~8.4\,GHz with the NASA Deep Space Network~(DSN) antenna, DSS-43. The observations and data reduction procedures are described in Section~\ref{Section:Observations}. In Section~\ref{Section:Results}, we provide measurements of the magnetar's profile shape, flux density, and radio spectrum. We also carry out a detailed single pulse analysis at 8.4\,GHz and study the morphology of individual pulses from the magnetar. We discuss and summarize our results in Section~\ref{Section:Discussion}. In this section, we consider the implication of our results on scattering through the~ISM toward the~GC. We also compare the emission properties of the GC~magnetar to other magnetars and high magnetic field pulsars. Lastly, we describe the similarities between the single pulse emission from this magnetar and fast radio bursts~(FRBs).


\section{Observations}
\label{Section:Observations}

High frequency radio observations of \text{PSR~J1745--2900} were carried out during four separate epochs between 2015~July~30 and 2016~August~20 using the NASA~DSN 70\,m antenna~(DSS-43) in Tidbinbilla, Australia. A detailed list of these radio observations is provided in Table~\ref{Table:RadioObservations}. Simultaneous dual circular polarization \mbox{$S$-band} and \mbox{$X$-band} data, centered at 2.3~and~8.4\,GHz, were recorded during each epoch with a time sampling of 512\,$\mu$s. The data were channelized, with a frequency spacing of 1\,MHz, in a digital polyphase filterbank with 96~and~480\,MHz of bandwidth at \mbox{$S$-band} and \mbox{$X$-band}, respectively. Polarimetric measurements are not provided since data from a polarimetry calibrator was unavailable.

These observations were performed at elevation angles between~12$^{\circ}$ and~21$^{\circ}$, and the antenna gain was~$\sim$1\,K/Jy. The total system temperature was calculated at \mbox{$S$/$X$-band} for each epoch using:
\begin{equation}
T_{\text{sys}}=T_{\text{rec}}+T_{\text{atm}}+T_{\text{GC}},
\label{Equation:TsysTotal}
\end{equation}
where $T_{\text{rec}}$ is the receiver noise temperature, $T_{\text{atm}}$ is the atmospheric contribution, and $T_{\text{GC}}$ is the contribution from the~GC. The atmospheric component was determined from the elevation angle, atmospheric optical depth, and atmospheric temperature during each epoch. In Table~\ref{Table:RadioObservations}, we list the sum of the instrumental and atmospheric components of the system temperature for each epoch at \mbox{$S$/$X$-band}, where we have assumed 15\%~uncertainties on these values. We modeled $T_{\text{GC}}$ using the following empirical relationship derived by~\citet{Rajwade2017} from calibrated continuum maps of the~GC~\citep{Law2008}:
\begin{equation}
T_{\text{GC}}(\nu)=568\left(\frac{\nu}{\text{GHz}}\right)^{-1.13}\text{K},
\label{Equation:TsysGC}
\end{equation}
where $\nu$ denotes the observing frequency. At the \mbox{$S$/$X$-band} central frequencies, the GC adds 227/52\,K to the system temperature, giving an average system temperature of 262(3)/78(2)\,K.


\begin{deluxetable*}{clccccc}
	\footnotesize
	\tablecaption{Radio Observations of \text{PSR~J1745--2900}}
	\tablewidth{0pt}
	\tablehead{
		\colhead{Epoch} &
		\colhead{Date$^{\mathrm{a}}$} &
		\colhead{Time$^{\mathrm{a}}$} &
		\colhead{Date$^{\mathrm{b}}$} &
		\colhead{Duration} &
		\colhead{$T_{\text{rec}}+T_{\text{atm}}$$^{\mathrm{c}}$} &
		\colhead{$T_{\text{rec}}+T_{\text{atm}}$$^{\mathrm{d}}$} \\
		\colhead{} &
		\colhead{} &
		\colhead{(hh:mm:ss)} &
		\colhead{(MJD)} &
		\colhead{(hr)} &
		\colhead{(K)} &
		\colhead{(K)}
	}
	\startdata
	1 & 2015 Jul 30 & 16:15:00 & 57233.67708 & 1.2 & 34\,$\pm$\,5 & 24\,$\pm$\,4 \\
	2 & 2015 Aug 15 & 15:25:22 & 57249.64262 & 1.3 & 35\,$\pm$\,5 & 25\,$\pm$\,4 \\
	3 & 2016 Apr 01 & 12:35:42 & 57479.52479 & 0.4 & 38\,$\pm$\,6 & 29\,$\pm$\,4 \\
	4 & 2016 Aug 20 & 15:26:28 & 57620.64338 & 1.0 & 36\,$\pm$\,5 & 27\,$\pm$\,4
	\enddata
	\tablecomments{\\
		$^{\mathrm{a}}$ Start time of the observation (UTC). \\
		$^{\mathrm{b}}$ Start time of the observation. \\
		$^{\mathrm{c}}$ Sum of the instrumental and atmospheric components of the system temperature at \mbox{$S$-band}. \\
		$^{\mathrm{d}}$ Sum of the instrumental and atmospheric components of the system temperature at \mbox{$X$-band}.}
	\label{Table:RadioObservations}
\end{deluxetable*}


\subsection{Data Reduction}
\label{Section:Data_Reduction}

The raw filterbank data are comprised of power spectral measurements across the band and can include spurious signals due to radio frequency interference~(RFI). The first step in the data reduction procedure was to remove data that were consistent with either narrowband or wideband~RFI. We searched the data using the \texttt{rfifind} tool from the \texttt{PRESTO}\footnote{See https://www.cv.nrao.edu/$\sim$sransom/presto.} pulsar search package~\citep{Ransom2001}, which produced a mask for filtering out data identified as~RFI and resulted in the removal of less than~3\% of the data from each epoch.

Next, we flattened the bandpass response and removed low frequency variations in the baseline of each frequency channel by subtracting the moving average from each data point, which was calculated using 10\,s of data around each time sample. The sample times were corrected to the solar system barycenter using the \texttt{TEMPO}\footnote{See http://tempo.sourceforge.net.} timing analysis software, and the data were then incoherently dedispersed at the magnetar's nominal~DM of~1778\,pc\,cm$^{\text{--3}}$.


\section{Results}
\label{Section:Results}

\subsection{Average Pulse Profiles}
\label{Section:Average_Pulse_Profiles}

\begin{figure*}[t]
	\centering
	\begin{tabular}{ccc}
		
		\subfigure
		{
			\includegraphics[trim=0cm 0cm 0cm 0cm, clip=false, scale=0.45, angle=0]{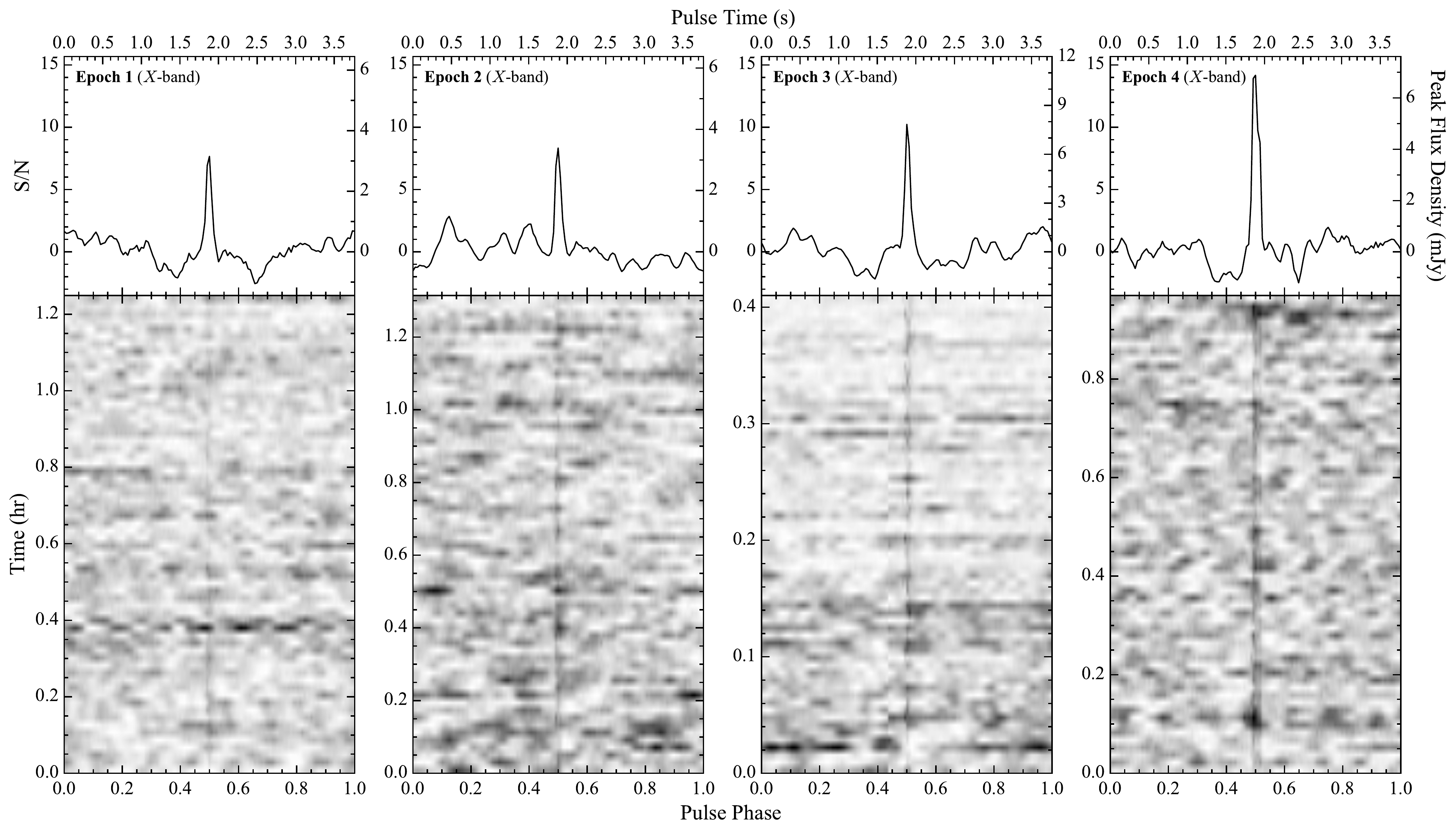}
			\label{Figure:Figure1a}
		}
		
		\\
		
		\subfigure
		{
			\includegraphics[trim=0cm 0cm 0cm 0cm, clip=false, scale=0.45, angle=0]{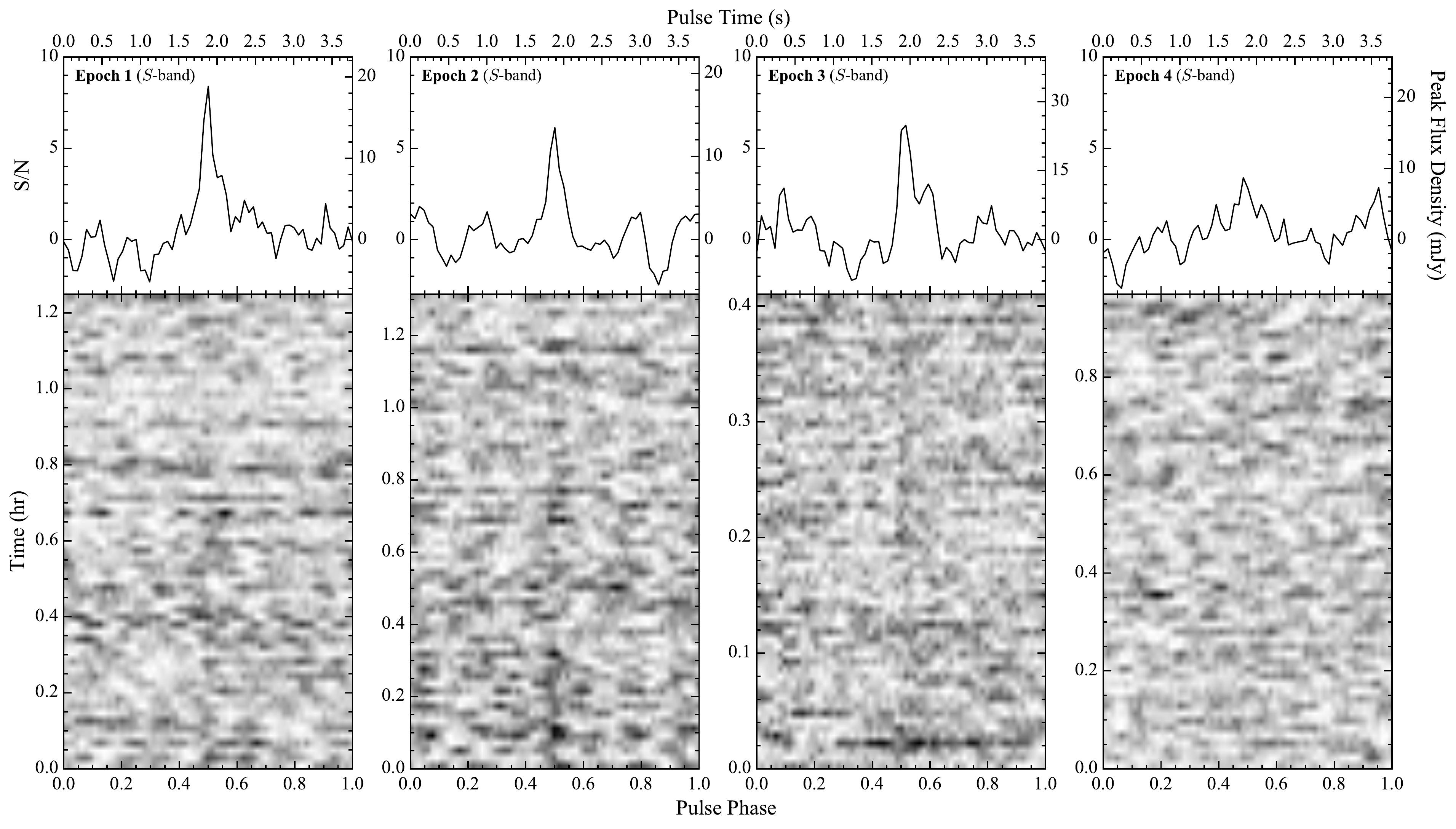}
			\label{Figure:Figure1b}
		}
		
	\end{tabular}
	
	\caption{Average pulse profiles of \text{PSR~J1745--2900} at (top row)~\mbox{$X$-band} and (bottom row)~\mbox{$S$-band} during epochs~\text{1--4} after combining data from both circular polarizations in quadrature. The data were folded on the barycentric period measurements given in Table~\ref{Table:PeriodMeasurements}. The top panels show the integrated pulse profiles using 64/128~phase bins at \mbox{$S$/$X$-band}, and the bottom panels show the strength of the pulsations as a function of phase and time, where darker bins correspond to stronger pulsed emission.}
	\label{Figure:Figure1}
\end{figure*}

A blind search for pulsations was performed between 3.6~and~3.9\,s using the \texttt{PRESTO} pulsar search package. Barycentric period measurements are provided in Table~\ref{Table:PeriodMeasurements} and were derived from the \mbox{$X$-band} data, where the pulsations were strongest. Average \mbox{$S$-band} and \mbox{$X$-band} pulse profiles, shown in Figure~\ref{Figure:Figure1}, were obtained after applying barycentric corrections, dedispersing at the magnetar's nominal~DM, and folding the data on the barycentric periods given in Table~\ref{Table:PeriodMeasurements}. These pulse profiles were produced by combining data from both circular polarizations in quadrature. The top panels show the integrated pulse profiles in units of peak flux density and signal-to-noise ratio~(S/N). The~S/N was calculated by subtracting the \text{off-pulse} mean from the pulse profiles and dividing by the \text{off-pulse} root mean square~(RMS)~noise level, $\sigma_{\text{off}}$. The bottom panels show the strength of the pulsations as a function of time and pulse phase. The pulse profiles have been aligned such that the peak of the \mbox{$X$-band} pulse profile lies at the center of the pulse phase window.

The \mbox{$X$-band} pulse profiles in Figure~\ref{Figure:Figure1} display a narrow emission component during each epoch, and the \mbox{$S$-band} pulse profiles from epochs~1--3 show broader peaks that are nearly coincident in phase with the \mbox{$X$-band} peaks. \mbox{$S$-band} pulsations were only marginally detected during epoch~4. From Figure~\ref{Figure:Figure1}, we see that the pulsed emission was stronger at \mbox{$X$-band} compared to \mbox{$S$-band} during epochs~\text{2--4}, but epoch~1 showed slightly more significant pulsations at \mbox{$S$-band}. The pulsations also became noticably fainter toward the end of epoch~1, and we found that the pulsed emission was weaker in the right circular polarization~(RCP) channel compared to the left circular polarization~(LCP) channel during this particular epoch.


\begin{deluxetable*}{clccc}
	\footnotesize
	\tablecaption{Barycentric Period Measurements of \text{PSR~J1745--2900}}
	\tablewidth{0pt}
	\tablehead{
		\colhead{Epoch} &
		\colhead{$P$} &
		\colhead{$\dot{P}$} &
		\colhead{$T_{\text{ref}}$$^{\mathrm{a}}$} \\
		\colhead{} &
		\colhead{(s)} &
		\colhead{(s\,s$^{\text{--1}}$)} &
		\colhead{(MJD)}
	}
	\startdata
	1 & 3.76531(1) & $<$\,2\,$\times$\,10$^{\text{--8}}$ &  57233.682318131 \\
	2 & 3.765367(8) & $<$\,1\,$\times$\,10$^{\text{--8}}$ & 57249.646691855 \\
	3 & 3.76603(2) & $<$\,1\,$\times$\,10$^{\text{--7}}$ & 57479.527055687 \\
	4 & 3.76655(1) & $<$\,2\,$\times$\,10$^{\text{--8}}$ & 57620.646961446
	\enddata
	\tablecomments{Period measurements were derived from the barycentered \mbox{$X$-band} data. \\
			$^{\mathrm{a}}$ Barycentric reference time of period measurements.}
	\label{Table:PeriodMeasurements}
\end{deluxetable*}



\subsection{Mean Flux Densities and Spectral Indices}
\label{Section:Spectral_Index}

Measurements of the magnetar's mean flux density were calculated from the average \mbox{$S$-band} and \mbox{$X$-band} pulse profiles in Figure~\ref{Figure:Figure1} using the modified radiometer equation~\citep{Lorimer2012}:
\begin{equation}
S_{\nu}=\frac{\beta\,T_{\text{sys}}\,(A_{\text{pulse}}/N_{\text{total}})}{G\sqrt{\Delta\nu\,n_{p}\,T_{\text{obs}}}},
\label{Equation:MeanFluxDensity}
\end{equation}
where $\beta$~is a correction factor that accounts for system imperfections such as digitization of the signal, $T_{\text{sys}}$~is the effective system temperature given by Equation~(\ref{Equation:TsysTotal}), $A_{\text{pulse}}$~is the area under the pulse, $G$~is the telescope gain, $N_{\text{total}}$\,$=$\,$\sqrt{n_{\text{bin}}}\,\sigma_{\text{off}}$ is the total RMS~noise level of the profile, $n_{\text{bin}}$~is the total number of phase bins in the profile, $\Delta\nu$~is the observing bandwidth, $n_{p}$~is the number of polarizations, and $T_{\text{obs}}$~is the total observation time. Errors on the mean flux densities were derived from the uncertainties in the flux calibration parameters. In Table~\ref{Table:FluxDensitiesSpectralIndices}, we provide a list of mean flux density measurements at 2.3~and~8.4\,GHz for each epoch. An upper limit is given for the \mbox{$S$-band} mean flux density during epoch~4 since pulsations were only marginally detected.

The \mbox{$X$-band} mean flux densities measured on 2015~July~30 and August~15 were smaller by a factor of~$\sim$7.5 compared to measurements made roughly 5~months earlier by~\citet{Torne2017}. Observations performed on 2016~April~1 and August~20 indicate that the magnetar's \mbox{$X$-band} mean flux density more than doubled since 2015~August~15. The \mbox{$S$-band} mean flux density was noticably variable, particularly during epoch~4 when a significant decrease in pulsed emission strength was observed. This behavior is not unusual, as large changes in radio flux densities have also been observed from other magnetars on short timescales~(e.g.,~\citealt{Levin2012}).

The spectral index,~$\alpha$, was calculated for each epoch using our simultaneous mean flux density measurements at 2.3~and~8.4\,GHz, assuming a power-law relationship of the form $S_{\nu}$\,$\propto$\,$\nu^{\alpha}$. These spectral index measurements are listed in Table~\ref{Table:FluxDensitiesSpectralIndices}. A wide range of spectral index values have been reported from multifrequency radio observations of this magnetar~\citep{Eatough2013b, Shannon2013, Pennucci2015, Torne2015, Torne2017}. \citet{Torne2017}~measured a spectral index of $\alpha$\,$=$\,$+$0.4\,$\pm$\,0.2 from radio observations between 2.54~and~291\,GHz between 2015~March~4 and~9, approximately 5~months prior to our observations. However, the radio spectrum derived by~\citet{Torne2017} was considerably steeper between 2.54~and~8.35\,GHz. We performed a nonlinear least squares fit using their total average flux densities in this frequency range and found a spectral index of \mbox{$\alpha$\,$=$\,--0.6\,$\pm$\,0.2}. Our spectral index measurements (see~Table~\ref{Table:FluxDensitiesSpectralIndices}) indicate that the magnetar exhibited a significantly negative average spectral index of \mbox{$\langle\alpha\rangle$\,$=$\,--1.86\,$\pm$\,0.02} during epochs~\text{1--3} when its 8.4\,GHz profile was \text{single-peaked}. The spectral index flattened to $\alpha$\,$>$\,--1.12 during epoch~4 when the profile became \text{double-peaked}~(see~Figure~\ref{Figure:Figure2}). While our spectral index values suggest a much steeper spectrum than is typical for the other three known radio magnetars, which have nearly flat or inverted spectra~\citep{Camilo2006, Camilo2008, Lazaridis2008, Levin2010, Keith2011}, a comparably steep spectrum has previously been observed from this magnetar between 2~and~9\,GHz~\citep{Pennucci2015}.


\begin{deluxetable*}{ccccc}
	\footnotesize
	\tablecaption{Flux Densities and Spectral Indices of \text{PSR~J1745--2900}}
	\tablewidth{0pt}
	\tablehead{
		\colhead{Epoch} &
		\colhead{$S_{\text{2.3}}$$^{\mathrm{a}}$} &
		\colhead{$S_{\text{8.4}}$$^{\mathrm{b}}$} &
		\colhead{$\alpha$$^{\mathrm{c}}$} \\
		\colhead{} &
		\colhead{(mJy)} &
		\colhead{(mJy)} &
		\colhead{}
	}
	\startdata
	1 & 1.18\,$\pm$\,0.02 & 0.078\,$\pm$\,0.004 & --2.08\,$\pm$\,0.04 \\
	2 & 0.79\,$\pm$\,0.02 & 0.085\,$\pm$\,0.004 & --1.70\,$\pm$\,0.04 \\
	3 & 1.92\,$\pm$\,0.04 & 0.18\,$\pm$\,0.01 & --1.80\,$\pm$\,0.04 \\
	4 & $<$\,0.84 & 0.19\,$\pm$\,0.01 & $>$\,--1.12
	\enddata
	\tablecomments{\\
		$^{\mathrm{a}}$ Mean flux density at 2.3\,GHz. \\
		$^{\mathrm{b}}$ Mean flux density at 8.4\,GHz. \\
		$^{\mathrm{c}}$ Spectral index between 2.3~and~8.4\,GHz.}
	\label{Table:FluxDensitiesSpectralIndices}
\end{deluxetable*}


\subsection{Rotation-resolved Pulse Profiles}
\label{Section:Rotation_Resolved_Profiles}

\begin{figure*}[t]
	\centering
	\includegraphics[trim=0cm 0cm 0cm 0cm, clip=false, scale=0.45, angle=0]{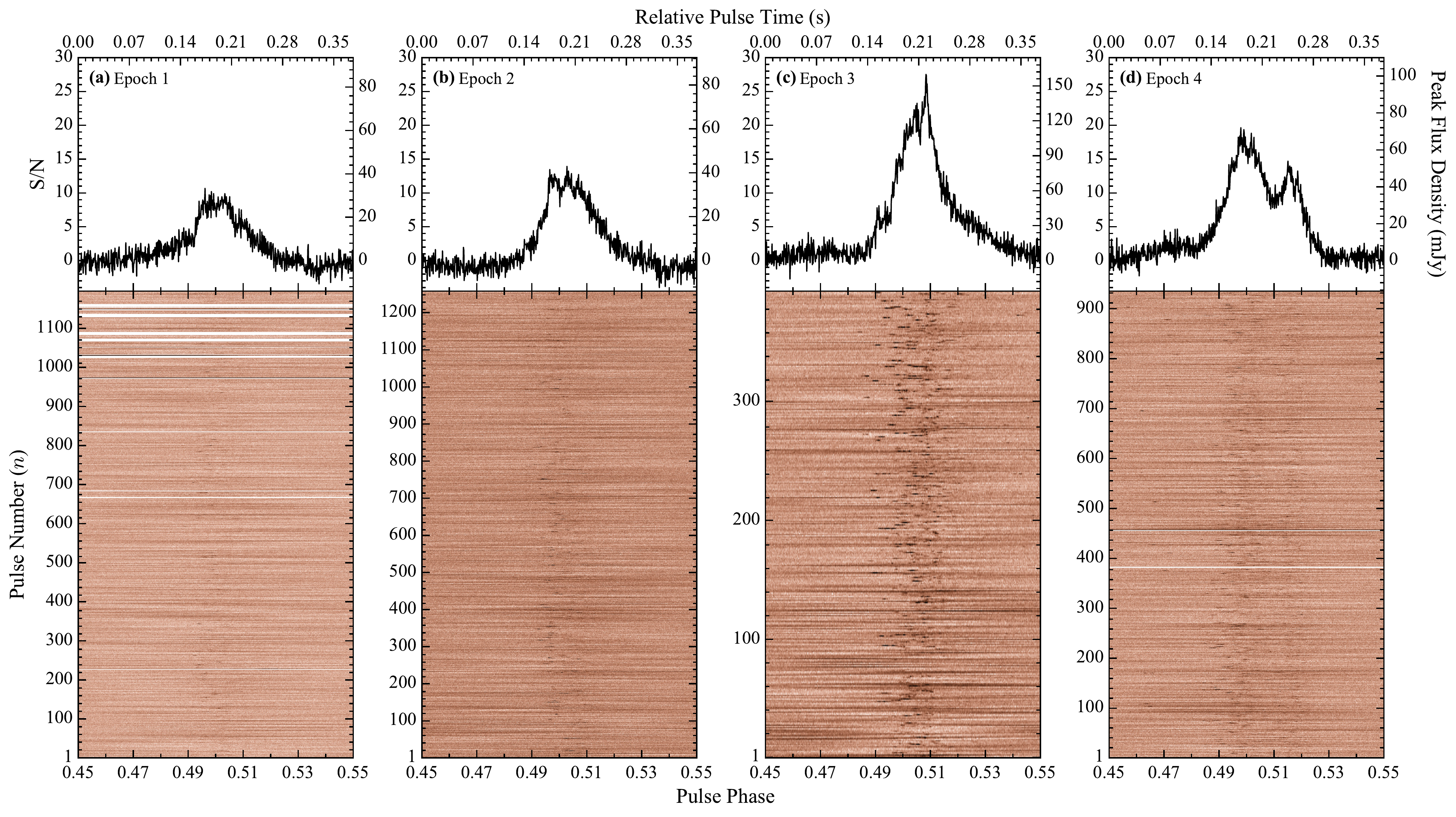}
	\caption{\text{Rotation-resolved} pulse profiles of \text{PSR~J1745--2900} at \mbox{$X$-band} during (a)~epoch~1, (b)~epoch~2, (c)~epoch~3, and (d)~epoch~4 after folding the data on the barycentric period measurements given in Table~\ref{Table:PeriodMeasurements} and combining data from both circular polarizations in quadrature. The data are shown with a time resolution of 512\,$\mu$s. The integrated profiles are displayed in the top panels, and the bottom panels show the distribution and relative strength of the single pulses as a function of pulse phase for each individual pulsar rotation, with darker bins signifying stronger emission. Pulse numbers are referenced with respect to the start of each observation.}
	\label{Figure:Figure2}
\end{figure*}

The \mbox{$X$-band} \text{rotation-resolved} pulse profiles in Figure~\ref{Figure:Figure2} were produced by folding the barycentered and dedispersed time series data on the barycentric periods given in Table~\ref{Table:PeriodMeasurements} and combining the data from both circular polarizations in quadrature. A time resolution of 512\,$\mu$s was used to define the spacing between neighboring phase bins. The bottom panels show the single pulse emission during each individual pulsar rotation as a function of pulse phase, and the integrated pulse profiles are shown in the top panels. In Figure~\ref{Figure:Figure2}, we show a restricted pulse phase interval~(0.45--0.55) around the \mbox{$X$-band} pulse profile peak from each epoch and reference pulse numbers with respect to the start of each observation. \mbox{$S$-band} \text{rotation-resolved} pulse profiles are not shown since the single pulse emission was significantly weaker at 2.3\,GHz.

The integrated profiles from epochs~\text{1--3}, shown in \text{Figures~\ref{Figure:Figure2}(a)--(c)}, exhibit a single feature with an approximately Gaussian shape, similar to previous observations near this frequency by~\citet{Spitler2014}. Finer substructure is also seen in the profiles, particularly during epoch~3 when the single pulse emission is brightest. Two main emission peaks are observed in the integrated profile from epoch~4, shown in Figure~\ref{Figure:Figure2}(d), with the secondary component originating from separate subpulses delayed by $\sim$65\,ms from the primary peak. \citet{Yan2015} also found subpulses that were coherent in phase over many rotations during observations with the~TMRT at 8.6\,GHz between 2014~June and October. We note that the shape of the average profile is mostly Gaussian during epoch~1 when the magnetar's radio spectrum is steepest and displays an additional component during epoch~4 after the spectrum has flattened (see~Table~\ref{Table:FluxDensitiesSpectralIndices}), which may suggest a link between the magnetar's radio spectrum and the structure of its pulsed emission.



\subsection{Single Pulse Analysis}
\label{Section:Single_Pulse_Analysis}

\subsubsection{Identification of Single Pulses}
\label{Section:Identification_Of_Single_Pulses}

We carried out a search for \mbox{$S$-band} and \mbox{$X$-band} single pulses from each epoch listed in Table~\ref{Table:RadioObservations}. In this paper, we focus primarily on \mbox{$X$-band} single pulses detected during epoch~3 since the single pulse emission was brightest during this epoch. The data were first barycentered and dedispersed at the magnetar's nominal~DM of~1778\,pc\,cm$^{\text{--3}}$ after masking bad data corrupted by~RFI and applying the bandpass and baseline corrections described in Section~\ref{Section:Data_Reduction}. The full time resolution time series data were then searched for single pulses using a Fourier domain matched filtering algorithm available through \texttt{PRESTO}, where the data were convolved with boxcar kernels of varying widths.

We used 54~boxcar templates with logarithmically spaced widths up to 2\,s, and events with S/N\,$\ge$\,5 were recorded for further analysis. If a single pulse candidate was detected with different boxcar widths from the same section of data, only the highest~S/N event was stored in the final list. The S/N of each single pulse candidate was calculated using:
\begin{equation}
\text{S/N}=\frac{\sum_{i}(f_{i}-\bar{\mu})}{\bar{\sigma}\sqrt{w}},
\label{Equation:SinglePulseSNR}
\end{equation}
where $f_{i}$~is the time series value in bin~$i$ of the boxcar function, $\bar{\mu}$ and $\bar{\sigma}$ are the local mean and RMS~noise after normalization, and $w$ is the boxcar width in number of bins. The time series data were detrended and normalized such that $\bar{\mu}$\,$\approx$\,0 and $\bar{\sigma}$\,$\approx$\,1. We note that the definition of~S/N in Equation~(\ref{Equation:SinglePulseSNR}) has the advantage of giving approximately the same result irrespective of how the input time series is downsampled, provided the pulse is still resolved~\citep{Deneva2016}.


\subsubsection{$X$-band Single Pulse Morphology}
\label{Section:Single_Pulse_Morphology}

\paragraph{Multiple Emission Components}
\label{Section:Multiple_Emission_Components}

An analysis was performed on the \mbox{$X$-band} single pulse events from epoch~3 that were both detected using the Fourier domain matched filtering algorithm described in Section~\ref{Section:Identification_Of_Single_Pulses} and showed resolvable dispersed pulses in their barycentered dynamic spectra. We measured the times of arrival~(ToAs) of the emission components comprising each single pulse event by incoherently dedispersing the barycentered dynamic spectra at the magnetar's nominal DM~of~1778\,pc\,cm$^{\text{--3}}$ and then searching for local maxima in the integrated single pulse profiles after smoothing the data by convolving the time series with a \text{one-dimensional} Gaussian kernel. The Gaussian kernel used in this procedure is given by:
\begin{equation}
K(t;\,\sigma)=\frac{1}{\sqrt{2\pi}\sigma}\text{exp}\left(-\frac{t^{2}}{2\sigma^{2}}\right),
\label{Equation:GaussianKernel}
\end{equation}
where $\sigma$ is the scale of the Gaussian kernel and $t$ corresponds to the sample time in the time series. A modest Gaussian kernel scale of 819\,$\mu$s was used to smooth the data, which did not hinder our ability to distinguish between narrow, closely spaced peaks. Individual emission components were identified as events displaying a dispersed feature in their dynamic spectrum along with a simultaneous peak in their integrated single pulse profile.

\begin{figure*}[t]
	\centering
	\begin{tabular}{ccc}
		
		\subfigure
		{
			\includegraphics[trim=0cm 0cm 0cm 0cm, clip=false, scale=0.36, angle=0]{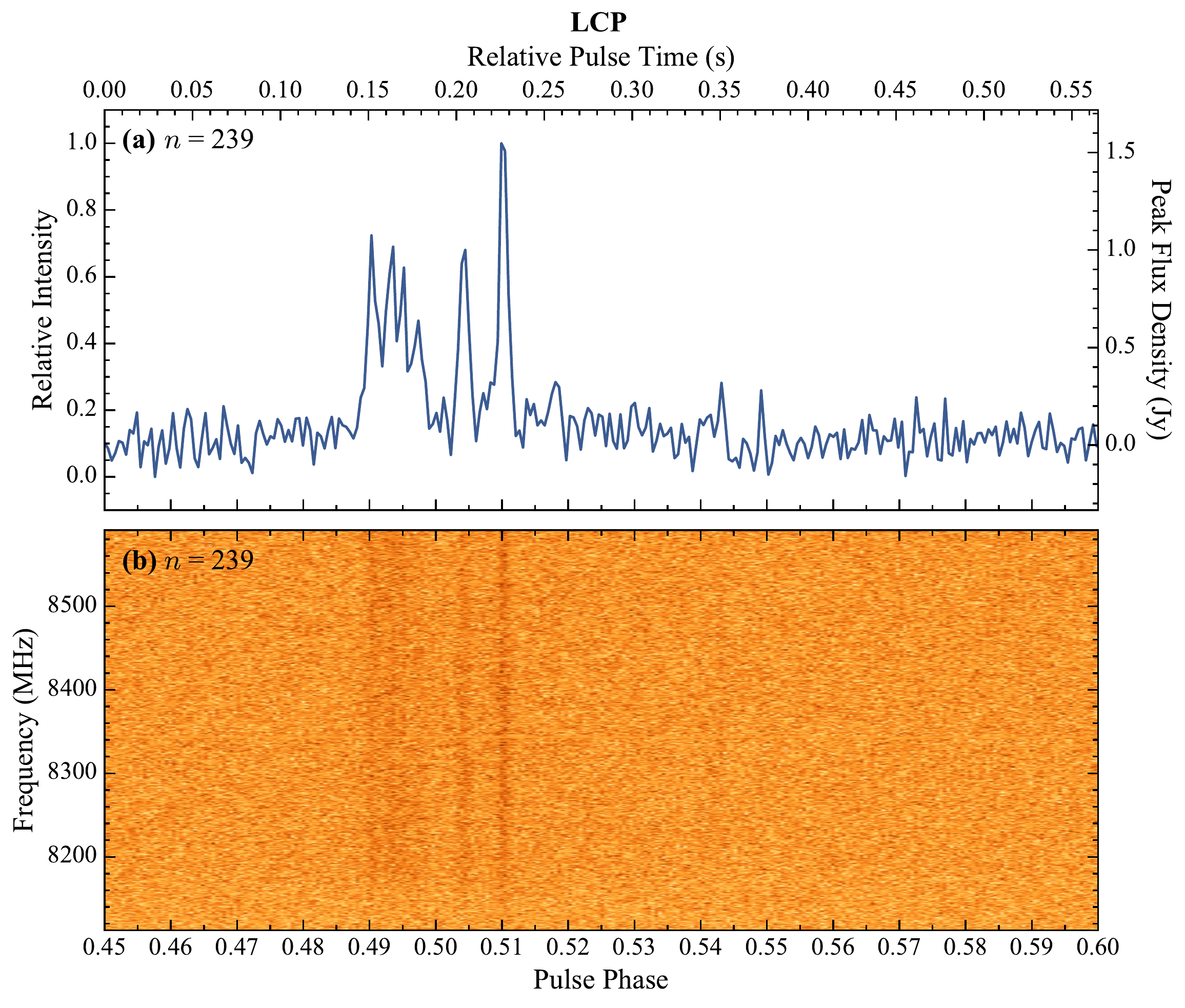}
			\label{Figure:Figure3a}
		}
		
		&
		
		\subfigure
		{
			\includegraphics[trim=0cm 0cm 0cm 0cm, clip=false, scale=0.36, angle=0]{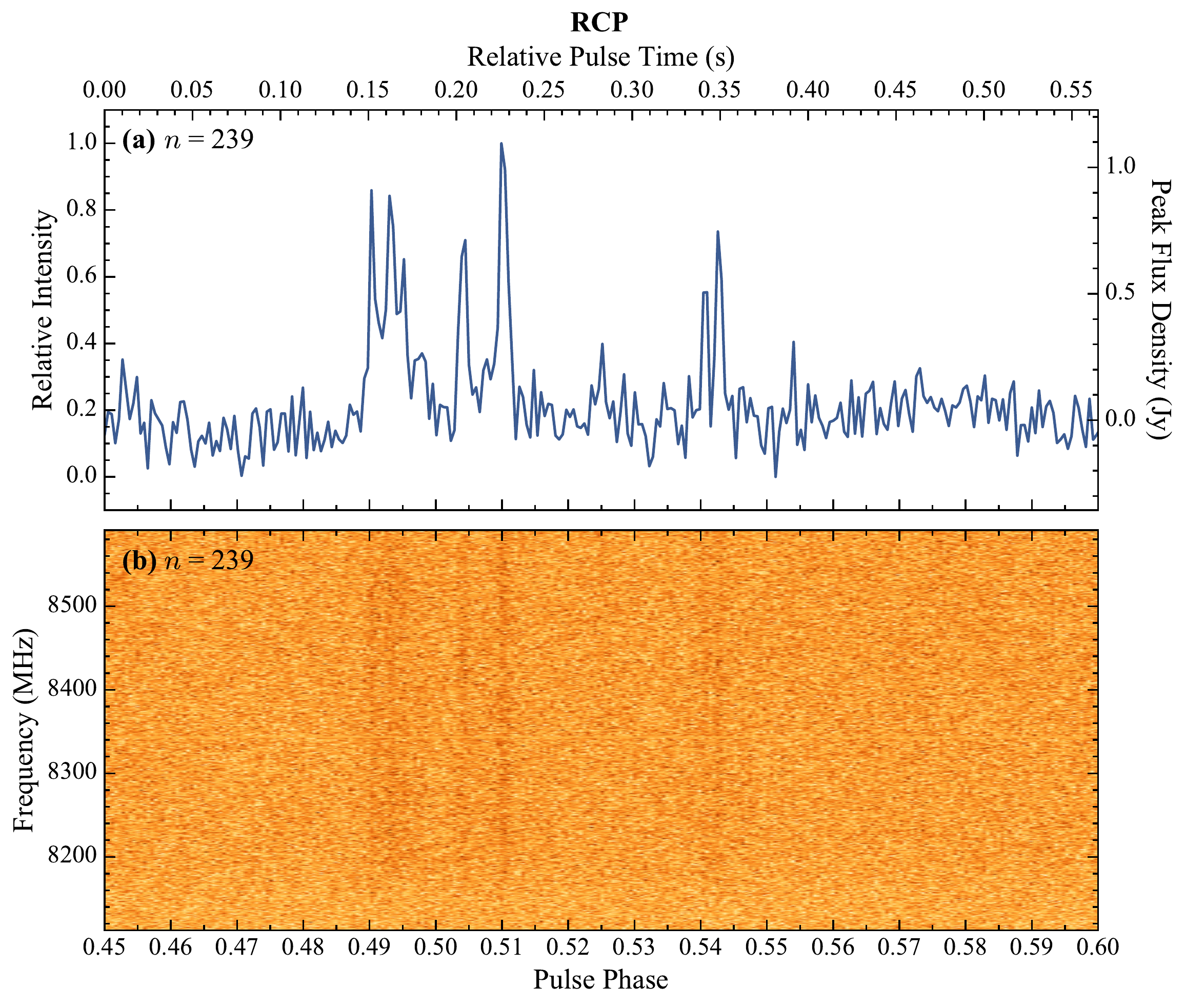}
			\label{Figure:Figure3b}
		}
		
		\\
		
		\subfigure
		{
			\includegraphics[trim=0cm 0cm 0cm 0cm, clip=false, scale=0.36, angle=0]{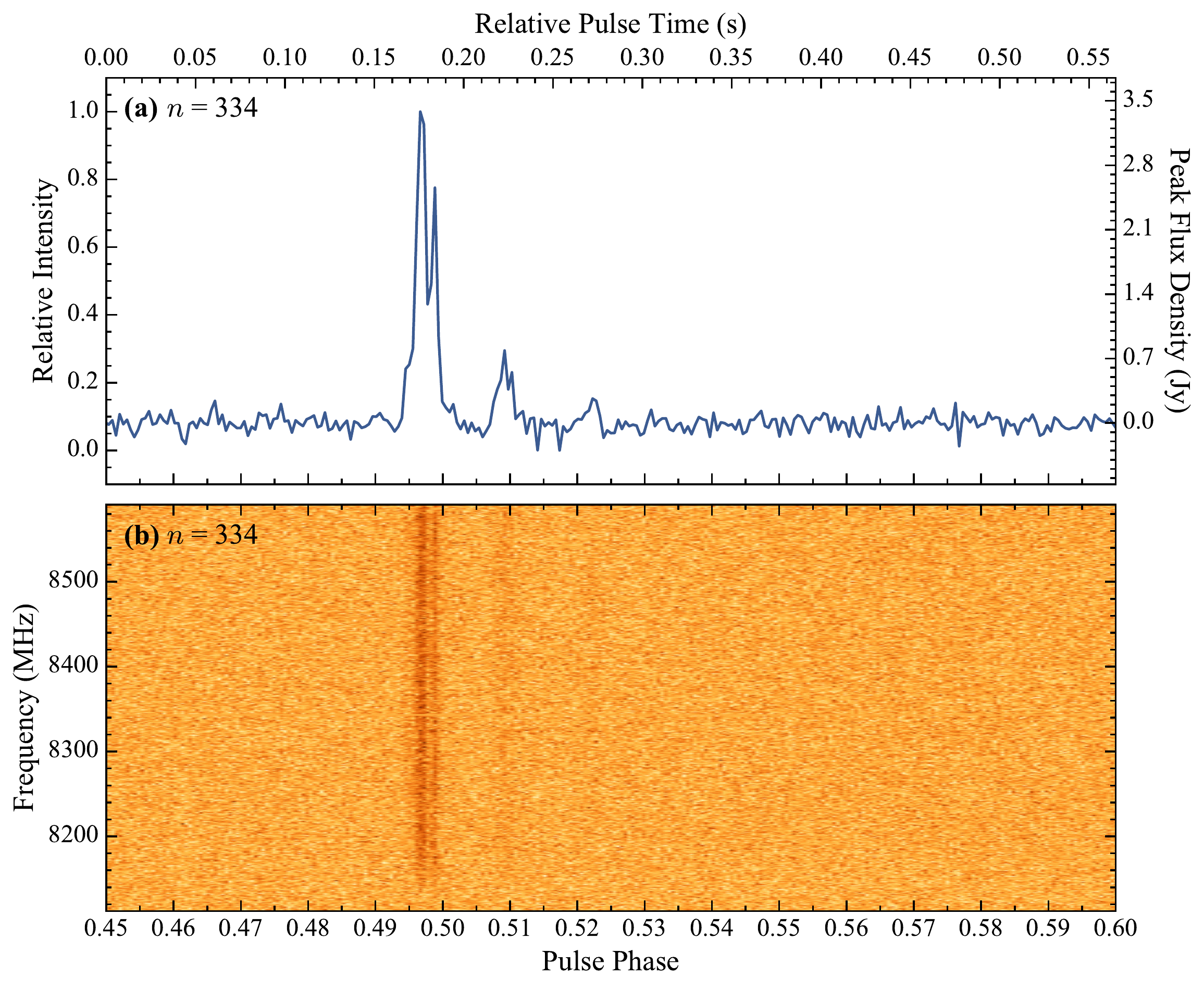}
			\label{Figure:Figure3c}
		}
		
		&
		
		\subfigure
		{
			\includegraphics[trim=0cm 0cm 0cm 0cm, clip=false, scale=0.36, angle=0]{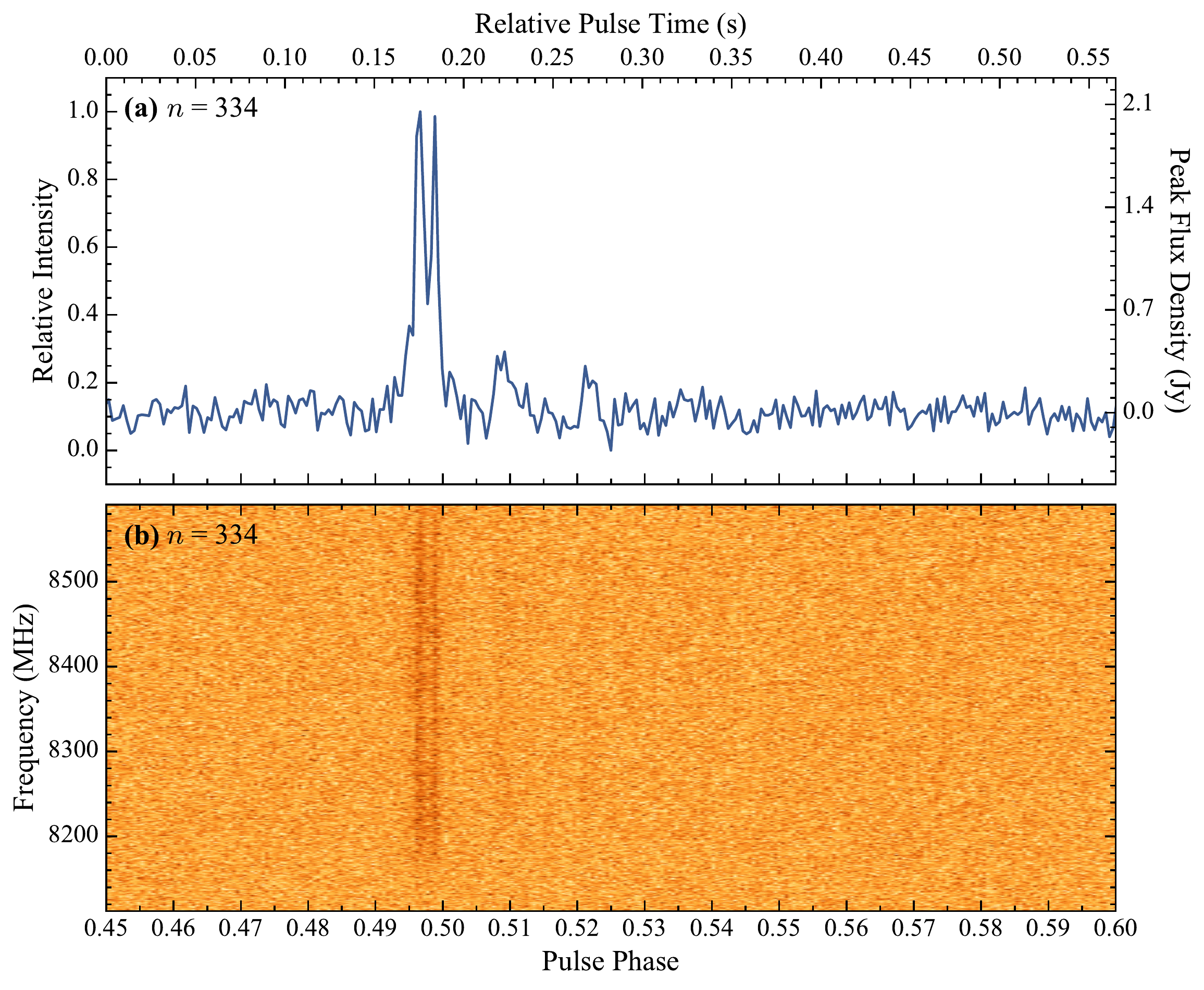}
			\label{Figure:Figure3d}
		}
		
		\\
		
		\subfigure
		{
			\includegraphics[trim=0cm 0cm 0cm 0cm, clip=false, scale=0.36, angle=0]{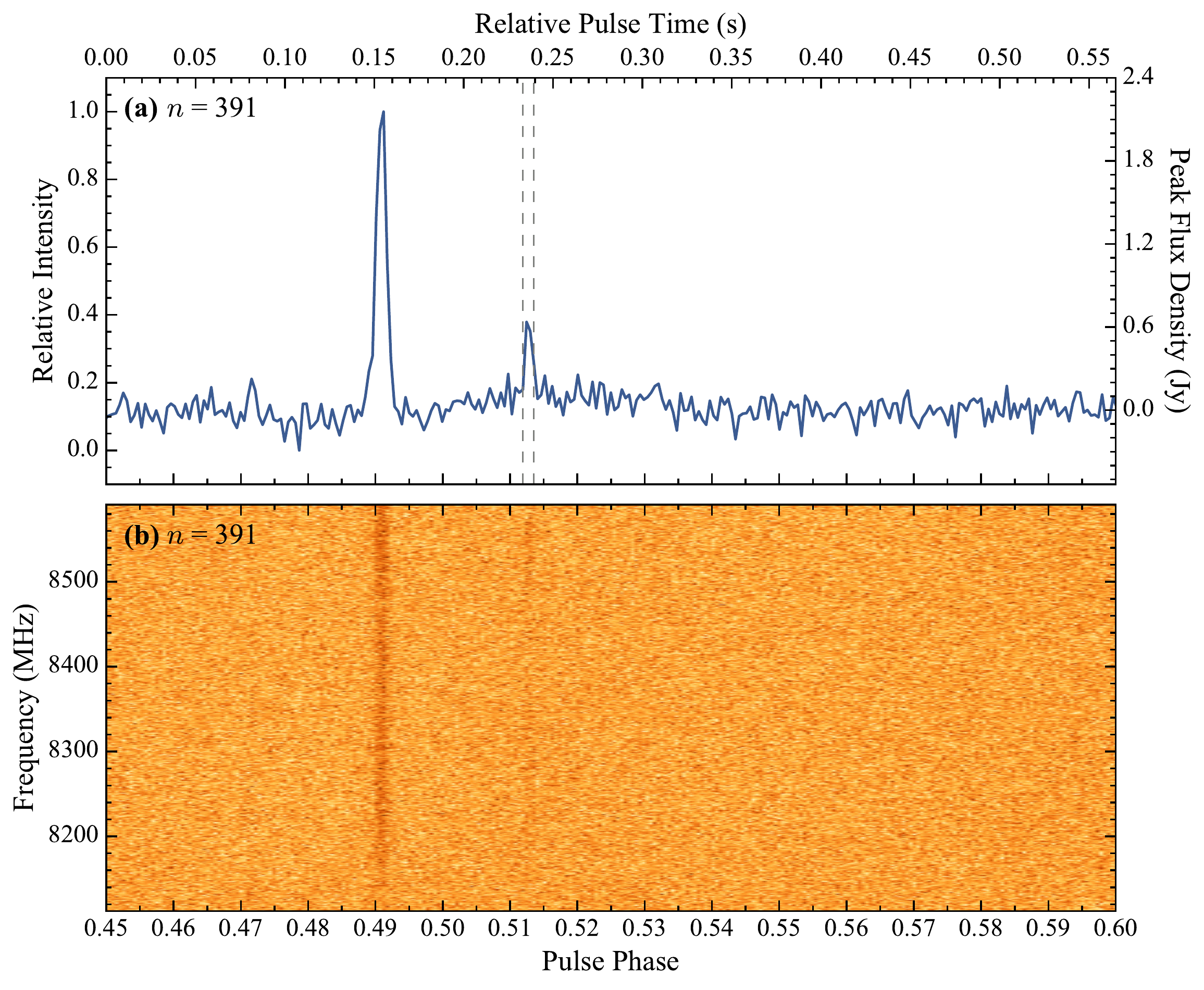}
			\label{Figure:Figure3e}
		}
		
		&
		
		\subfigure
		{
			\includegraphics[trim=0cm 0cm 0cm 0cm, clip=false, scale=0.36, angle=0]{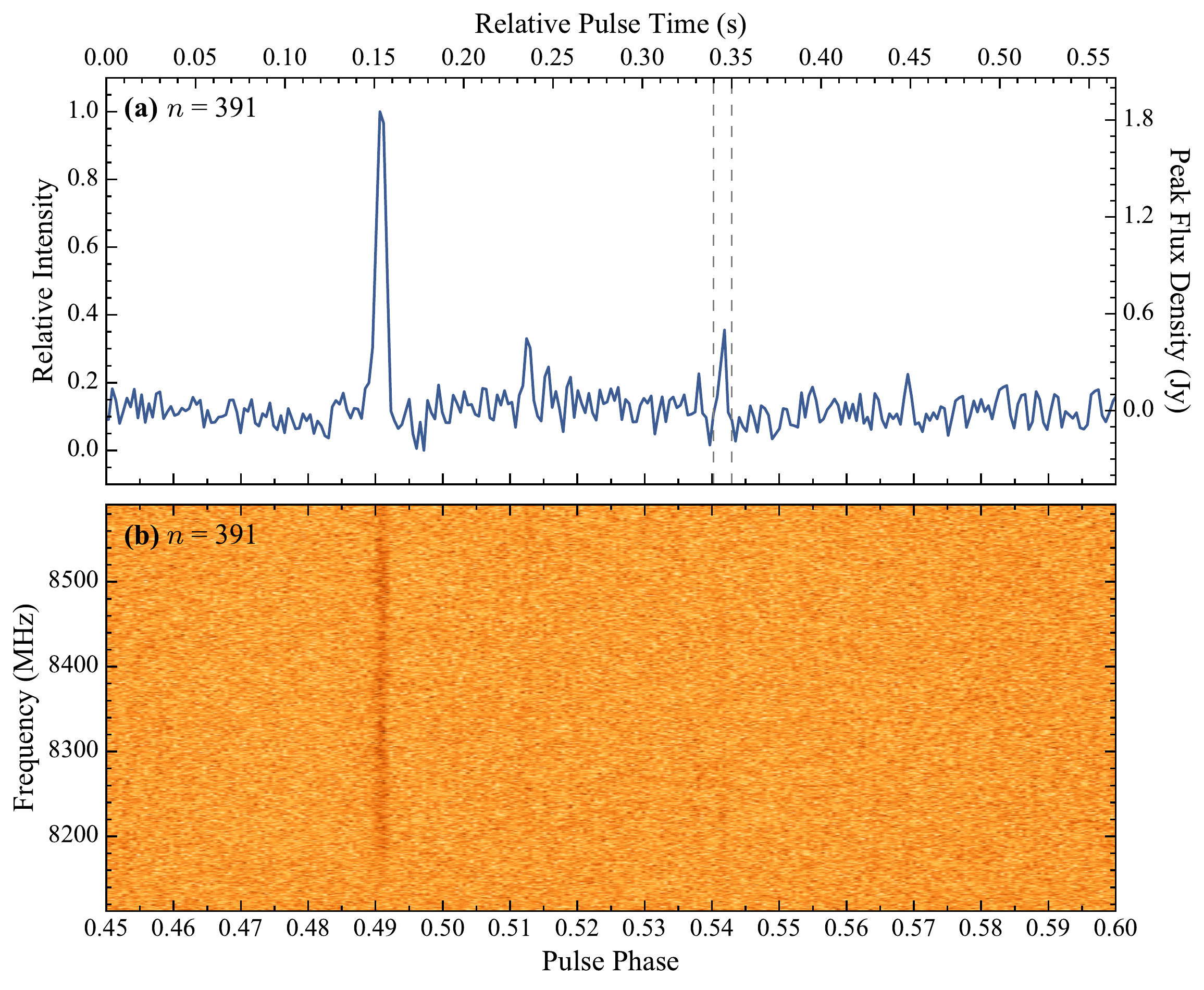}
			\label{Figure:Figure3f}
		}
		
	\end{tabular}
	
	\caption{Examples of bright \mbox{$X$-band} single pulse events displaying multiple emission components during pulse cycles (top~row)~$n$\,$=$\,239, (middle~row)~$n$\,$=$\,334, and (bottom~row)~$n$\,$=$\,391 of epoch~3. The plots in the left and right columns show detections of the single pulses in the left circular polarization~(LCP) and right circular polarization~(RCP) channels, respectively. We show the (a)~integrated single pulse profiles and (b)~dynamic spectra dedispersed using a DM~of~1778\,pc\,cm$^{\text{--3}}$ from both polarizations with a time resolution of 2\,ms. During pulse cycle~$n$\,$=$\,239, the emission components at later pulse phases are significantly detected in the RCP~channel, but are only marginally detected in the LCP~data. The two dominant emission components detected during pulse cycle~$n$\,$=$\,334 are separated by $\sim$6.8\,ms and $\sim$8.6\,ms in the~LCP and~RCP data, respectively. The secondary emission components detected during pulse cycle~$n$\,$=$\,391 show gaps across the frequency band in the dynamic spectra, but no frequency structure is observed in the primary component's emission. In Figure~\ref{Figure:Figure5}, we show the frequency structure of the secondary components indicated by the dashed vertical lines.}
	\label{Figure:Figure3}
\end{figure*}

The structure and number of \mbox{$X$-band} single pulse emission components varied significantly between consecutive pulsar rotations~(e.g.,~Figure~\ref{Figure:Figure2}). These changes were observed on timescales shorter than the magnetar's 3.77\,s rotation period. An example is shown in the top row of Figure~\ref{Figure:Figure3} from pulse cycle $n$\,$=$\,239 of epoch~3, where at least six distinct emission components can be resolved. While the overall structure of this particular single pulse is similar in the~LCP and RCP~channels, the emission components at later pulse phases are detected more strongly in the RCP~data. Measurements performed near this epoch at 8.35\,GHz with the Effelsberg telescope indicate that the magnetar likely had a high linear polarization fraction~\citep{Desvignes2018}. This suggests that some of the magnetar's emission components may be more polarized than others.

Other single pulse events contained fewer emission components. The middle row of Figure~\ref{Figure:Figure3} shows a single pulse event detected during pulse cycle $n$\,$=$\,334 of epoch~3 with four independent emission components in the~LCP and RCP~channels. The two brightest components are separated by $\sim$6.8\,ms and $\sim$8.6\,ms in the~LCP and RCP~data, respectively. Another example from pulse cycle $n$\,$=$\,391 of epoch~3 is provided in the bottom row of Figure~\ref{Figure:Figure3}, which shows two emission components in the LCP~data and three components in the RCP~data.

\begin{figure}[t]
	\centering
	\includegraphics[trim=0cm 0cm 0cm 0cm, clip=false, scale=0.485, angle=0]{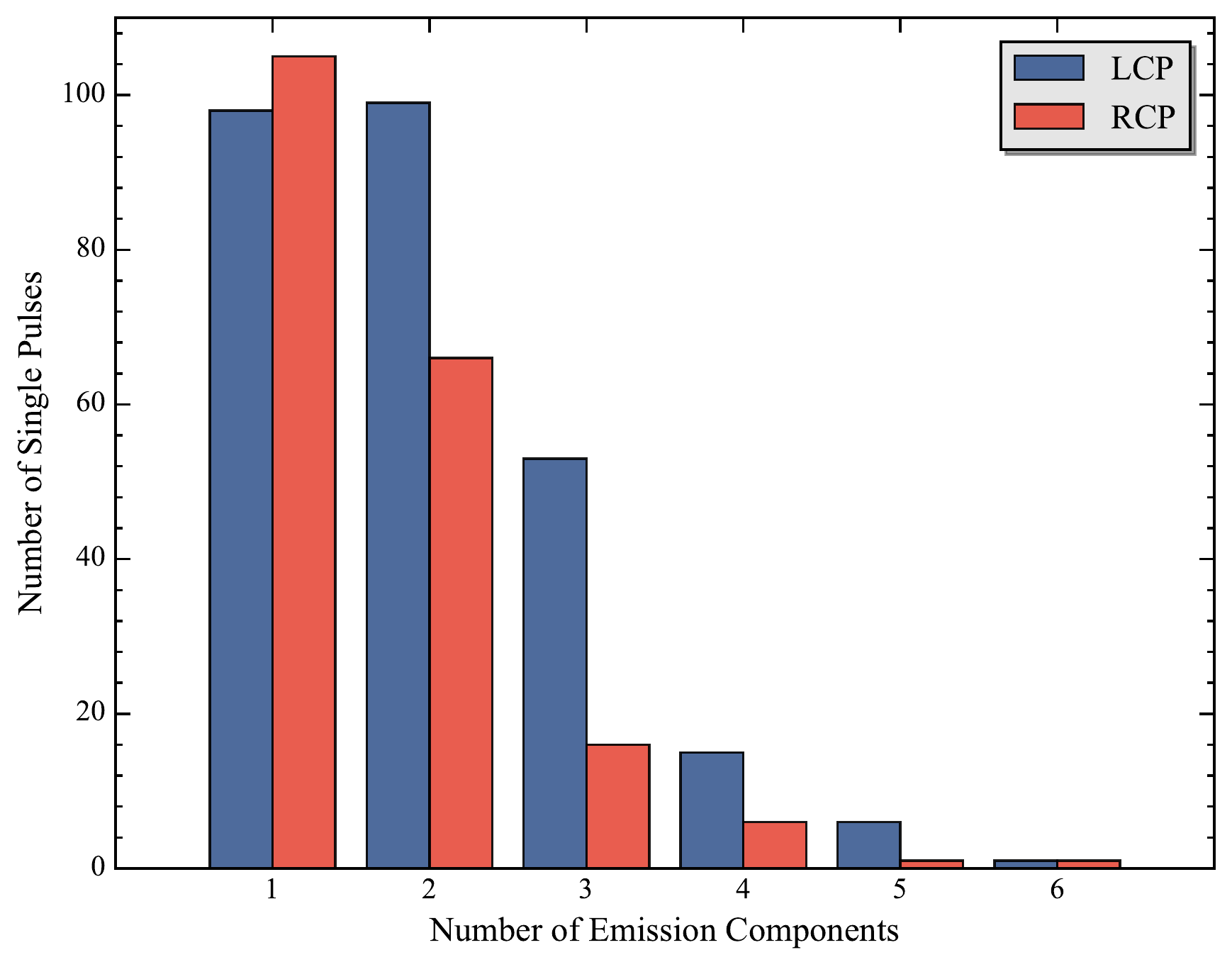}
	\caption{Number of \mbox{$X$-band} single pulse emission components detected during epoch~3 in the (blue)~left circular polarization~(LCP) and (red)~right circular polarization~(RCP) channels.}
	\label{Figure:Figure4}
\end{figure}

Using the threshold criteria described in Section~\ref{Section:Identification_Of_Single_Pulses}, single pulse emission components were significantly detected in at least one of the polarization channels during 72\% of the pulse cycles in epoch~3 and were identified in the LCP/RCP~data during 69\%/50\% of the pulse cycles. Faint emission components were often seen in many of the single pulses, but at a much lower significance level. In Figure~\ref{Figure:Figure4}, we show the distribution of the number of significantly detected emission components during these pulse cycles. More than 72\%/87\% of the single pulses in the LCP/RCP~data contained either one or two distinct emission components. The number of single pulses with either one or two emission components was approximately equal in the LCP~data, and 59\% more single pulses in the RCP~channel were found to have one emission component compared to the number of events with two components.

Most of the \mbox{$X$-band} single pulses detected during epochs~1~and~2 displayed only one emission component, whereas the single pulses from epochs~3 and~4 showed multiple emission components. Single pulses with multiple emission components have also been previously detected at 8.7\,GHz with the phased~VLA~\citep{Bower2014} and at 8.6\,GHz with the TMRT~\citep{Yan2015}. In both studies, the number of emission components and structure of the single pulses were found to be variable between pulsar rotations.



\paragraph{Frequency Structure in Emission Components}
\label{Section:Frequency_Structure}

Many of the \mbox{$X$-band} single pulse emission components from epoch~3 displayed frequency structure in their dynamic spectra. These events were characterized by a disappearance or weakening of the radio emission over subintervals of the frequency bandwidth. The typical scale of these frequency features was approximately 100\,MHz in extent, and the location of these features often varied between components in a single pulse cycle. The flux densities of these components also varied by factors of \mbox{$\sim$2--10} over the affected frequency ranges. From a visual inspection of the data, we find that approximately 50\% of the pulse cycles exhibited this behavior, and the LCP/RCP~data showed these features during 40\%/30\% of the pulse cycles. While these effects were often more pronounced in one of the polarization channels, approximately 20\% of the pulse cycles displayed events with these features in both channels simultaneously. This behavior was usually observed in the fainter emission components, but frequency structure was also sometimes seen in the primary component. These effects are not instrumental in origin since only a subset of the components were affected during a given pulse cycle.

\begin{figure}[t]
	\centering
	\includegraphics[trim=0cm 0cm 0cm 0cm, clip=false, scale=0.39, angle=0]{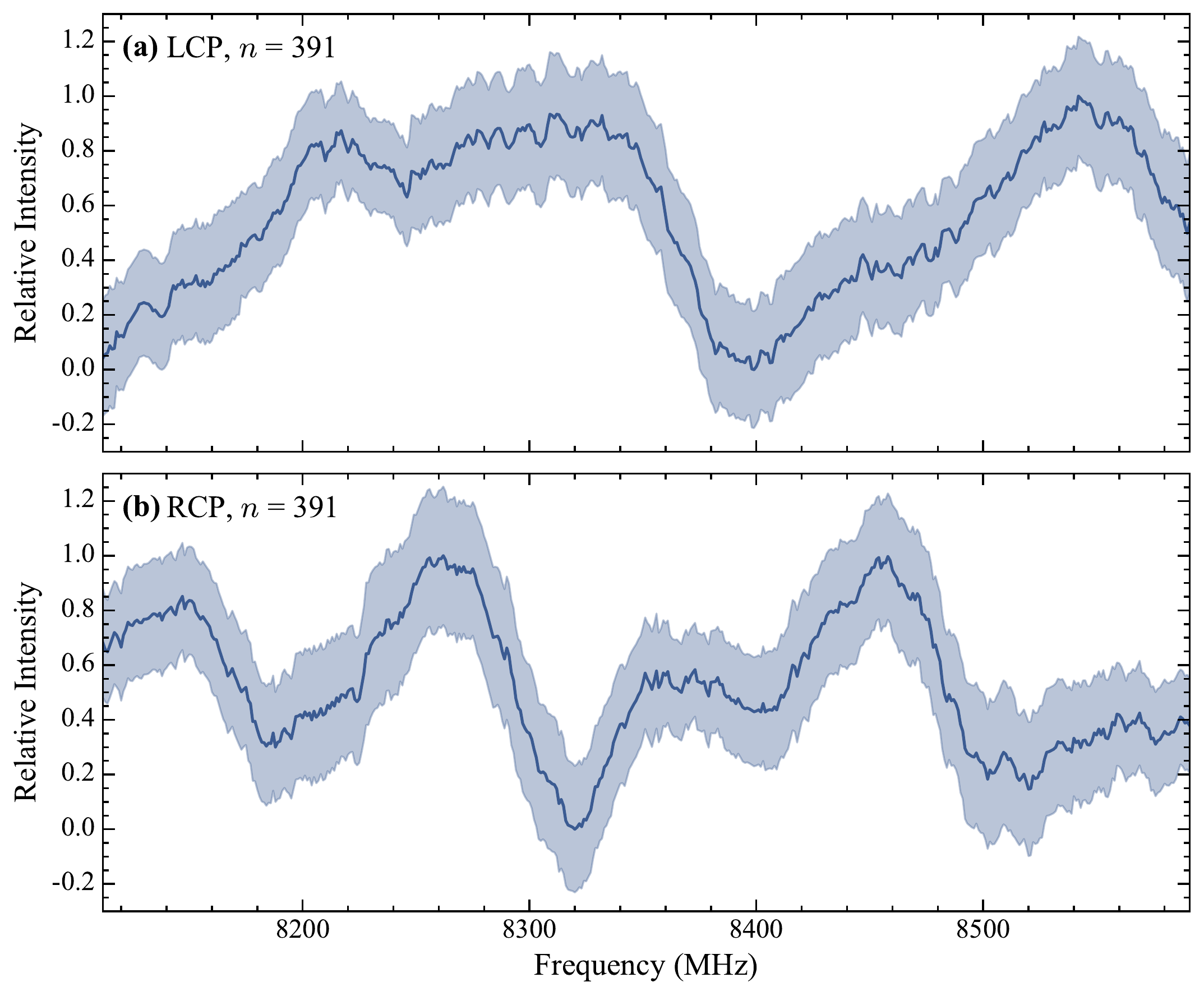}
	\caption{Examples of frequency structure in the secondary emission components of the \mbox{$X$-band} single pulse event during pulse cycle $n$\,$=$\,391 of epoch~3. The frequency structure in the (a)~left circular polarization~(LCP) and (b)~right circular polarization~(RCP) channels correspond to the secondary components labeled by the dashed vertical lines in the bottom row of Figure~\ref{Figure:Figure3}. The frequency response of the components is smoothed using a \text{one-dimensional} Gaussian kernel with $\sigma$\,$=$\,25\,MHz, and thus neighboring points are correlated. The blue shaded regions indicate the standard errors on the data points. The secondary component in the LCP~data shows a frequency gap centered at $\sim$8.4\,GHz spanning $\sim$100\,MHz. The frequency structure of the secondary component from the RCP~data is more complex and shows a gap near $\sim$8.3\,GHz.}
	\label{Figure:Figure5}
\end{figure}

We show an example single pulse from pulse cycle $n$\,$=$\,391 of epoch~3 in the bottom row of Figure~\ref{Figure:Figure3}, where gaps in the radio emission were observed in the secondary emission components. We selected two secondary components displaying this behavior, which we indicate with dashed vertical lines in Figure~\ref{Figure:Figure3}, and show their frequency structure in Figure~\ref{Figure:Figure5}. We used a \text{one-dimensional} Gaussian kernel with $\sigma$\,$=$\,25\,MHz, defined in Equation~(\ref{Equation:GaussianKernel}), to smooth the frequency response of these secondary components. The uncertainty associated with each data point was calculated from the standard error and is indicated by the blue shaded regions in Figure~\ref{Figure:Figure5}. The selected emission component in the LCP~data displays a frequency gap centered at $\sim$8.4\,GHz spanning $\sim$100\,MHz. The emission component from the RCP~data exhibits more complex frequency structure with a gap near $\sim$8.3\,GHz.

Next, we investigate whether the observed frequency structure could be produced by interstellar scintillation. Assuming a scattering timescale of $\tau_{d}$\,$\approx$\,$\nu^{\text{--4}}$, where $\nu$ is the observing frequency in~GHz, we estimate the diffractive interstellar scintillation bandwidth using:
\begin{equation}
\Delta\nu_{\text{DISS}}=\frac{C_{1}}{2\pi\tau_{d}},
\label{Equation:Scintillation}
\end{equation}
where $C_{1}$\,$=$\,1.16 for a uniform medium with a Kolmogorov wavenumber spectrum~\citep{Cordes1998}. At the \mbox{$X$-band} observing frequency of 8.4\,GHz, a scattering timescale of $\tau_{d}$\,$\approx$\,0.2\,ms corresponds to a predicted scintillation bandwidth of $\Delta\nu_{\text{DISS}}$\,$\approx$\,1\,kHz, which is well below the frequency scale associated with these features. The scintillation bandwidth decreases if we adopt a larger scattering timescale, such as the $\sim$7\,ms characteristic single pulse broadening timescale reported in Section~\ref{Section:Pulse_Broadening}. Therefore, interstellar scintillation cannot explain the frequency structure observed in the emission components. This behavior is likely caused by propagation effects in the magnetar's local environment, but may also be intrinsic to the magnetar.



\paragraph{Flux Density Distribution of Emission Components}
\label{Section:Flux_Density_Distribution}

We performed a statistical analysis of the distribution of peak flux densities from the \mbox{$X$-band} single pulse emission components detected during epoch~3. For each emission component, the peak~S/N was calculated by dividing the barycentered, integrated single pulse profiles by the \text{off-pulse} RMS~noise level after subtracting the \text{off-pulse} mean. The peak flux density of each emission component was determined from~\citep{McLaughlin2003}:
\begin{equation}
S_{\text{peak}}=\frac{\beta\,T_{\text{sys}}\,(\text{S/N})_{\text{peak}}}{G\,\sigma_{\text{off}}\sqrt{\Delta\nu\,n_{p}\,t_{\text{peak}}}},
\end{equation}
where $(\text{S/N})_{\text{peak}}$ is the peak~S/N of the emission component and $t_{\text{peak}}$ denotes the integration time at the peak.

\begin{figure}[t]
	\centering
	\includegraphics[trim=0cm 0cm 0cm 0cm, clip=false, scale=0.485, angle=0]{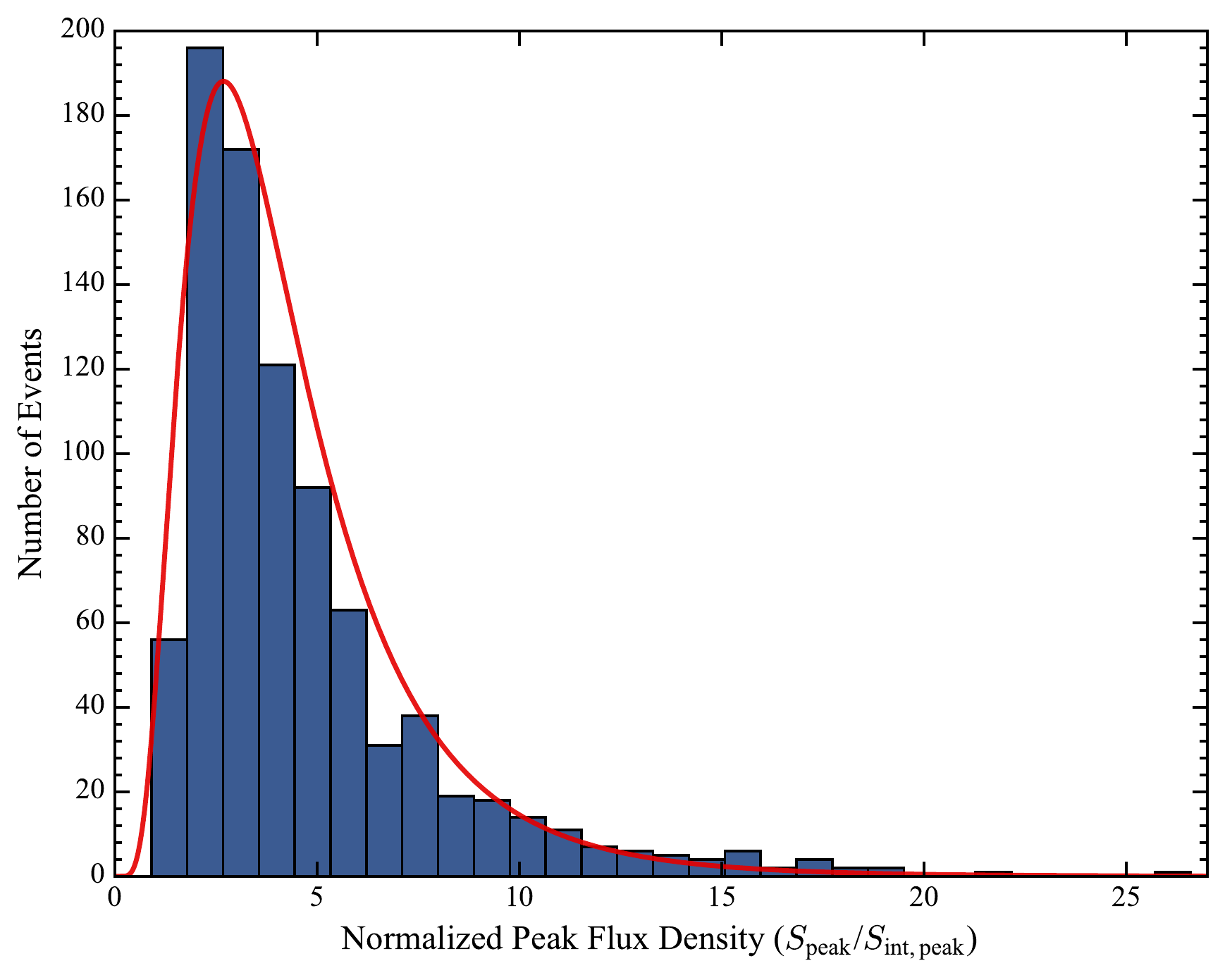}
	\caption{Distribution of peak flux densities from the single pulse emission components detected at \mbox{$X$-band} during epoch~3. The flux densities are normalized by $S_{\text{int,peak}}$\,$=$\,0.16\,Jy, the peak flux density from the integrated \text{rotation-resolved} profile in Figure~\ref{Figure:Figure2}(c). The \text{best-fit} \text{log-normal} distribution is overlaid in red. A high flux tail is observed in the distribution due to bright emission components with $S_{\text{peak}}$\,$\gtrsim$\,15\,$S_{\text{int,peak}}$.}
	\label{Figure:Figure6}
\end{figure}

Following the analyses in~\citet{Lynch2015} and~\citet{Yan2015}, the peak flux density of each single pulse emission component was normalized by the peak flux density of the integrated \text{rotation-resolved} profile in Figure~\ref{Figure:Figure2}(c), $S_{\text{int,peak}}$\,$=$\,0.16\,Jy. A histogram of the distribution of normalized peak flux densities for all 871~emission components is shown in Figure~\ref{Figure:Figure6}. We investigated whether the normalized peak flux densities could be characterized by a \text{log-normal} distribution with a probability density function~(PDF) given by:
\small
\begin{equation}
P_{\text{LN}}\left(x=\frac{S_{\text{peak}}}{S_{\text{int,peak}}}\right)=\frac{1}{\sqrt{2\pi}\sigma_{\text{LN}}x}\text{exp}\left[-\frac{\left(\ln x-\mu_{\text{LN}}\right)^{2}}{2\sigma_{\text{LN}}^{2}}\right],
\label{Equation:LogNormalPDF}
\end{equation}
\normalsize
where $\mu_{\text{LN}}$ and $\sigma_{\text{LN}}$ are the mean and standard deviation of the distribution, respectively. A nonlinear least squares fit to the normalized peak flux densities using Equation~(\ref{Equation:LogNormalPDF}) gave a \text{best-fit} \text{log-normal} distribution with $\mu_{\text{LN}}$\,$=$\,1.33\,$\pm$\,0.03 and $\sigma_{\text{LN}}$\,$=$\,0.58\,$\pm$\,0.02, which is overlaid in red in Figure~\ref{Figure:Figure6}.

A Kolmogorov--Smirnov~(KS) test~\citep{Lilliefors1967} yielded a $p$-value of~0.044, which revealed that the normalized flux densities were marginally inconsistent with the fitted \text{log-normal} distribution. This is primarily due to the moderate number of bright emission components with $S_{\text{peak}}$\,$\gtrsim$\,15\,$S_{\text{int,peak}}$. These components form a high flux tail in the observed flux density distribution, which is underestimated by the derived \text{log-normal} distribution. After removing these events and repeating the KS~test, we obtained a $p$-value of~0.057 and \text{best-fit} \text{log-normal} mean and standard deviation values that were consistent with our previous fit. This indicates that emission components with $S_{\text{peak}}$\,$\lesssim$\,15\,$S_{\text{int,peak}}$ can be described by the \text{log-normal} distribution shown in red in Figure~\ref{Figure:Figure6}.

Our \text{best-fit} \text{log-normal} mean and standard deviation values are consistent to within 1$\sigma$ with the single pulse flux density distribution derived from measurements with the~TMRT at 8.6\,GHz on MJD~56836~\citep{Yan2015}. However, \citet{Yan2015} found that the distribution was consistent with \text{log-normal} and no high flux tail was observed. In contrast, a high flux tail was seen in the distribution of single pulse flux densities measured at 8.7\,GHz with the~GBT~\citep{Lynch2015}. Single pulse energy distributions at 1.4,~4.9,~and~8.35\,GHz from the radio magnetar \text{XTE~J1810--197} also showed \text{log-normal} behavior along with a high energy tail, which was modeled with a power law~\citep{Serylak2009}. Here, we fit a power law to the flux densities in the tail of the distribution and find a scaling exponent of $\Gamma$\,$=$\,--7\,$\pm$\,1 for events with $S_{\text{peak}}$\,$\ge$\,15\,$S_{\text{int,peak}}$.

Giant radio pulses are characterized by events with flux densities larger than ten times the mean flux level. We detected a total of 61~emission components with $S_{\text{peak}}$\,$\ge$\,10\,$S_{\text{int,peak}}$, which comprised 7\% of the events. Giant pulses were seen during 9\% of the pulse cycles, and 72\% of these events occurred during the second half of the observation. No correlation was found between the flux density and number of components in these bright events. Previous studies of single pulse flux densities at \mbox{$X$-band} also showed some evidence of giant single pulses from this magnetar, but at a much lower rate~\citep{Yan2015}.



\paragraph{Temporal Variability of Emission Components}
\label{Section:Temporal_Variability}

\begin{figure*}[t]
	\centering
	\includegraphics[trim=0cm 0cm 0cm 0cm, clip=false, scale=0.37, angle=0]{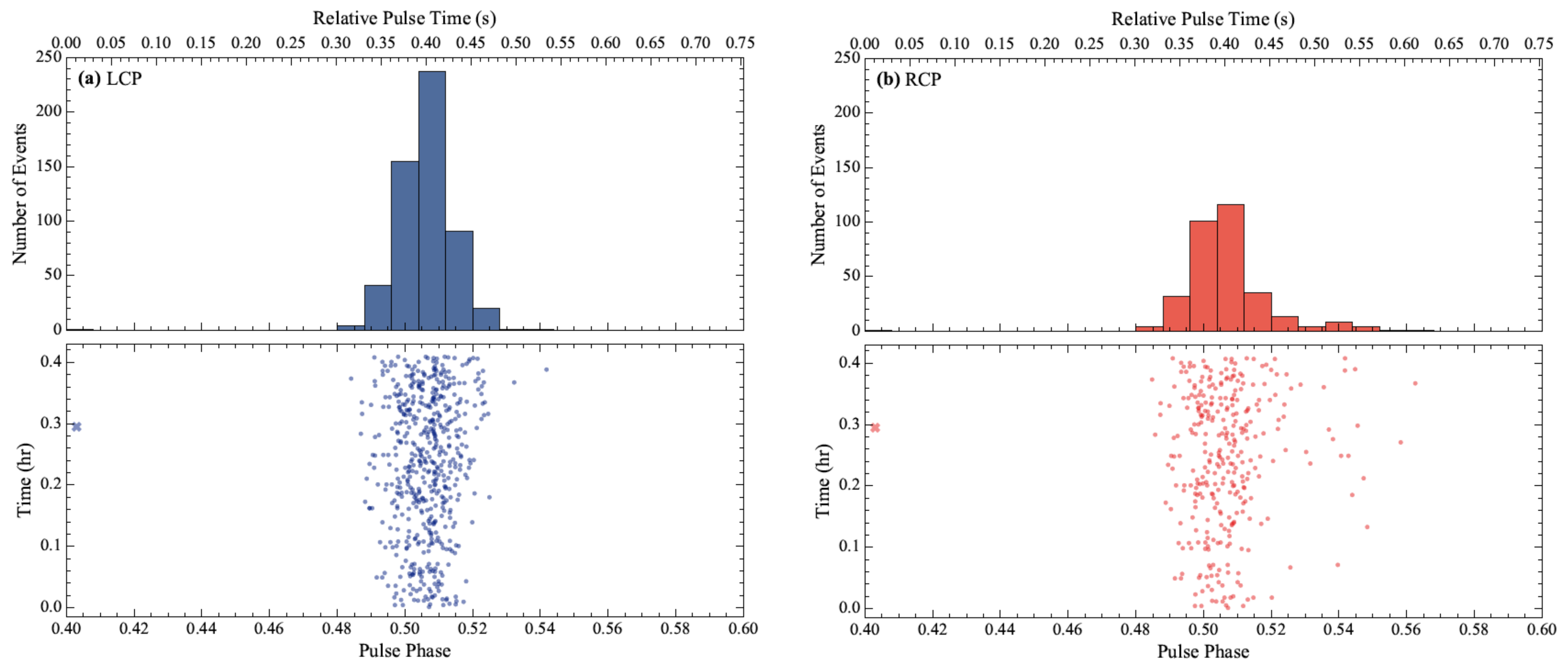}
	\caption{\mbox{$X$-band} single pulse emission components detected during epoch~3 in the (a)~left circular polarization~(LCP) and (b)~right circular polarization~(RCP) channels. Histograms of the number of events detected at each pulse phase are shown in the top panels, and the bottom panels show the phase distributions of the components from folding their times of arrival~(ToAs) on the barycentric period in Table~\ref{Table:PeriodMeasurements}. A bright single pulse, indicated with a cross, was detected earlier in pulse phase (near~$\sim$0.4) relative to the other events.}
	\label{Figure:Figure7}
\end{figure*}

The \mbox{$X$-band} single pulses from epoch~3 exhibited significant temporal variability between their emission components. To study this behavior, we folded the ToAs associated with the emission components on the barycentric period given in Table~\ref{Table:PeriodMeasurements}. The distribution of these events in pulse phase is shown in Figure~\ref{Figure:Figure7} for both polarization channels. A larger number of components were detected at later phases in the RCP~channel compared to the LCP~channel. This produced a tail in the phase distribution of events from the RCP~data, which can be seen in the top panel of Figure~\ref{Figure:Figure7}(b). No tail was observed in the phase distribution of the components from the LCP~data since emission components at later phases were generally not detected as strongly (e.g.,~top row of Figure~\ref{Figure:Figure3}).

A bright single pulse with one emission component was detected much earlier in pulse phase (near~$\sim$0.4) relative to the other events and is indicated by a cross in the bottom panels of Figure~\ref{Figure:Figure7}. Only two other single pulses were found at similarly anomalous phases during epoch~4. All of these pulses displayed a single, narrow emission component, and their observed pulse widths ranged between 1.5~and~3.1\,ms based on the boxcar widths used for detection in the matched filtering algorithm~(see Section~\ref{Section:Identification_Of_Single_Pulses}). This suggests that these pulses are either atypical for this magnetar or unrelated to the pulsar. Excluding this deviant event from our epoch~3 analysis, we measure pulse jitter values of $\sigma_{\text{LCP}}$\,$\approx$\,28\,ms and $\sigma_{\text{RCP}}$\,$\approx$\,44\,ms from the standard deviation of the emission component pulse phases in each polarization channel.

Approximately 41\%/38\% of the components in the LCP/RCP~data were detected at larger pulse phases than the peak phase of the integrated \text{rotation-resolved} profile in Figure~\ref{Figure:Figure2}(c). We also found that 29\%/41\% of the components in the LCP/RCP~channels were delayed by more than 30\,ms from the profile peak, which indicates that the emission components in the LCP~data are more tightly clustered in phase.

\begin{figure*}[t]
	\centering
	\begin{tabular}{ccc}
		
		\subfigure
		{
			\includegraphics[trim=0cm 0cm 0cm 0cm, clip=false, scale=0.35, angle=0]{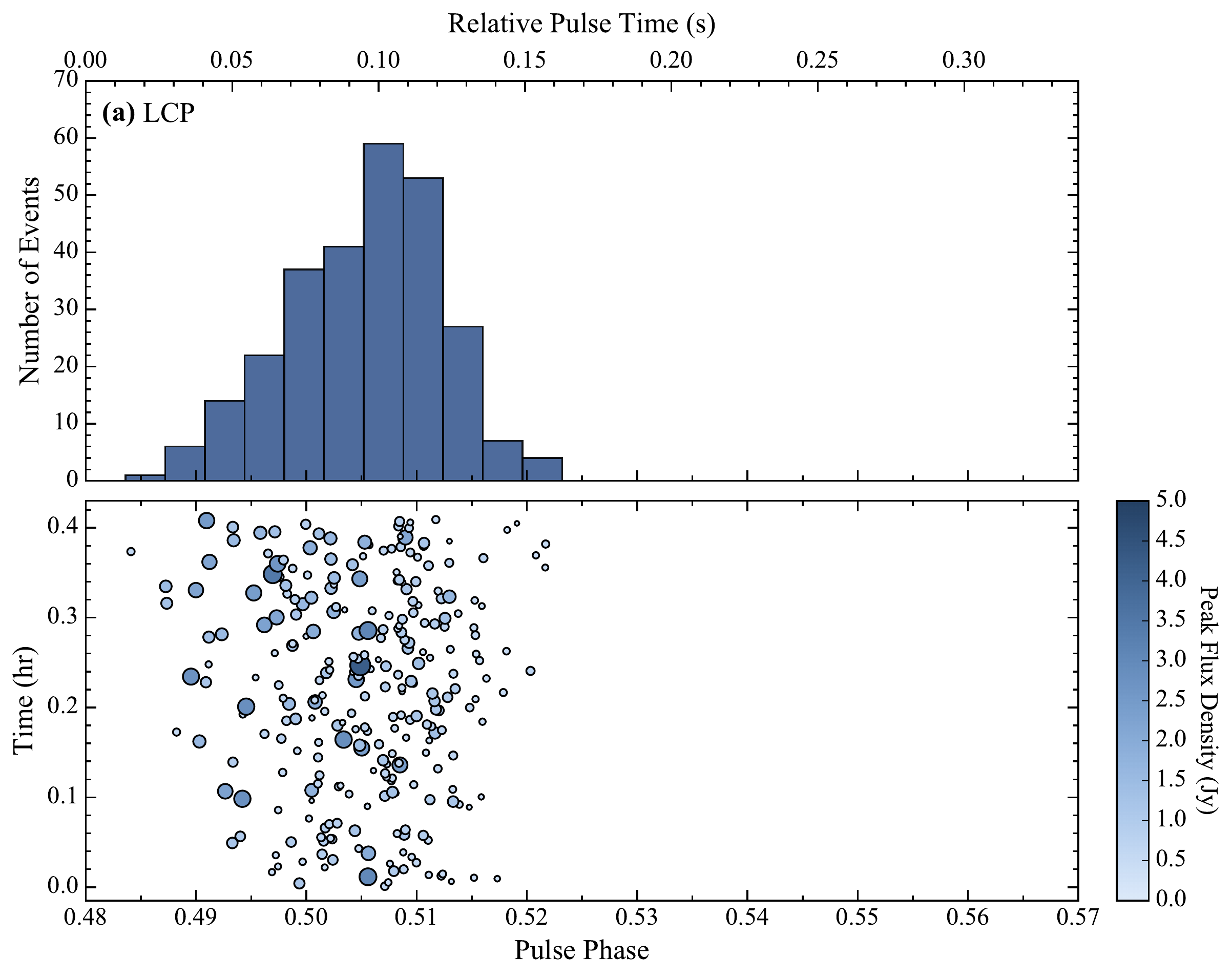}
			\label{Figure:Figure8a}
		}
		
		&
		
		\subfigure
		{
			\includegraphics[trim=0cm 0cm 0cm 0cm, clip=false, scale=0.35, angle=0]{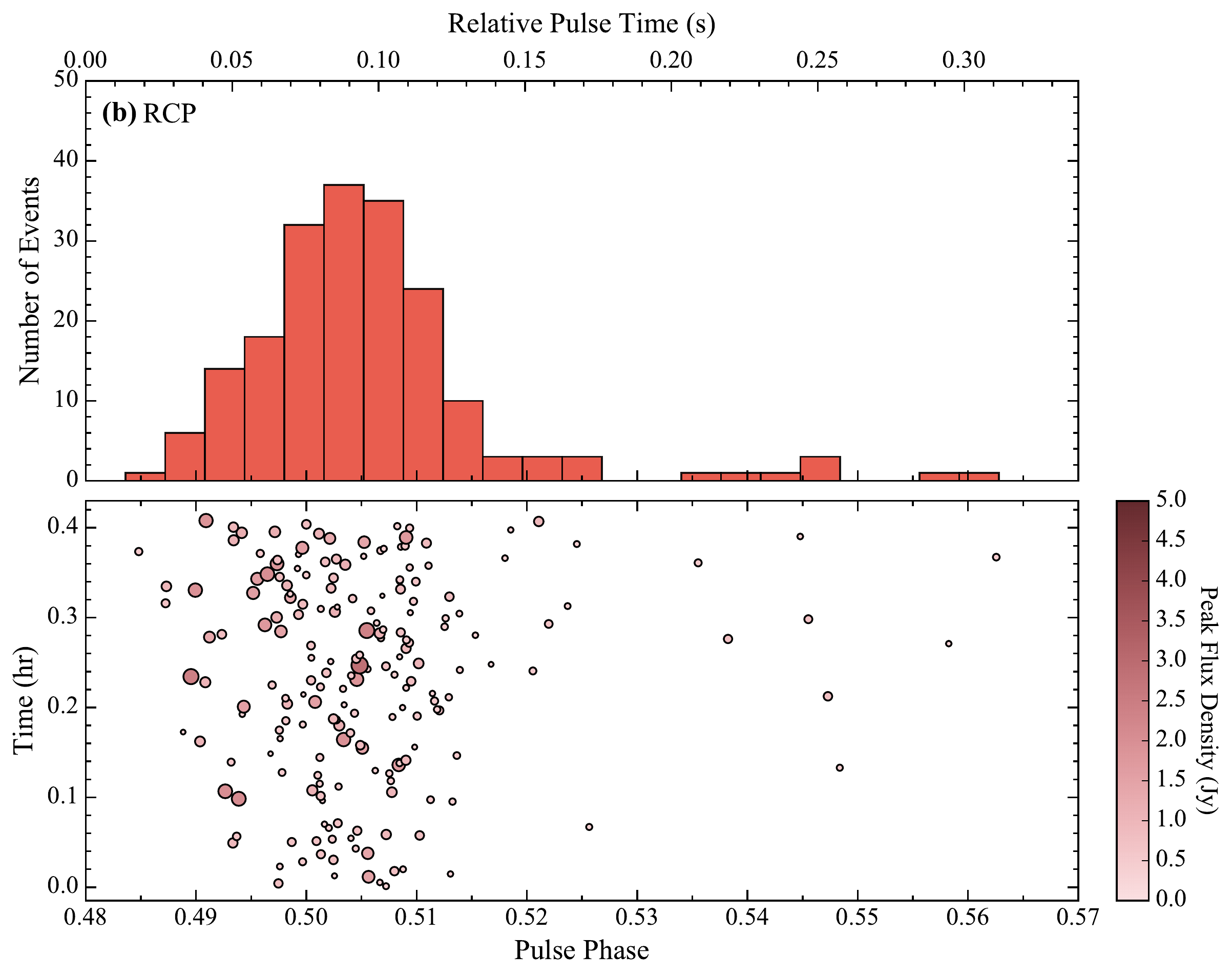}
			\label{Figure:Figure8b}
		}
		
	\end{tabular}
	
	\caption{Brightest \mbox{$X$-band} single pulse emission component detected during each pulsar rotation in Figure~\ref{Figure:Figure7}. Events exceeding the threshold criteria defined in Section~\ref{Section:Identification_Of_Single_Pulses} in the (a)~left circular polarization~(LCP) and (b)~right circular polarization~(RCP) channels are shown in blue and red, respectively. The top panels show histograms of the number of events at each pulse phase. Phase distributions of the components, determined from folding the times of arrival~(ToAs) on the barycentric period in Table~\ref{Table:PeriodMeasurements}, are shown in the bottom panels, where larger and darker circles correspond to events with larger peak flux densities. We excise the single pulse event near pulse phase~$\sim$0.4 from both polarization channels to show the distributions over a narrower phase range.}
	\label{Figure:Figure8}
\end{figure*}

In Figure~\ref{Figure:Figure8}, we show the pulse phase and peak flux density of the brightest emission component in each pulse cycle from Figure~\ref{Figure:Figure7}, where larger and darker circles in the bottom panels correspond to larger peak flux densities. A tail is again observed in the phase distribution of the components from the RCP~channel, and no tail is produced by components from the LCP~data. Pulse jitter values of $\sigma_{\text{LCP}}$\,$\approx$\,26\,ms and $\sigma_{\text{RCP}}$\,$\approx$\,43\,ms were found from the LCP and RCP~phase distributions shown in Figure~\ref{Figure:Figure8}, and these values are similar to those obtained from the distributions in Figure~\ref{Figure:Figure7}.

The relative time delays between the single pulse emission components also varied between pulsar rotations. During some pulse cycles, the single pulses showed a multicomponent structure with two bright components separated by $\lesssim$\,10\,ms, along with additional emission components with larger time delays. An example is shown in the middle row of Figure~\ref{Figure:Figure3}, where different time delays between the two brightest components were found in the two polarization channels. In other cases, only one dominant emission component was detected and the time delays between neighboring components were much larger~(e.g.,~bottom row of Figure~\ref{Figure:Figure3}).

\begin{figure*}[t]
	\centering
	\begin{tabular}{ccc}
		
		\subfigure
		{
			\includegraphics[trim=0cm 0cm 0cm 0cm, clip=false, scale=0.39, angle=0]{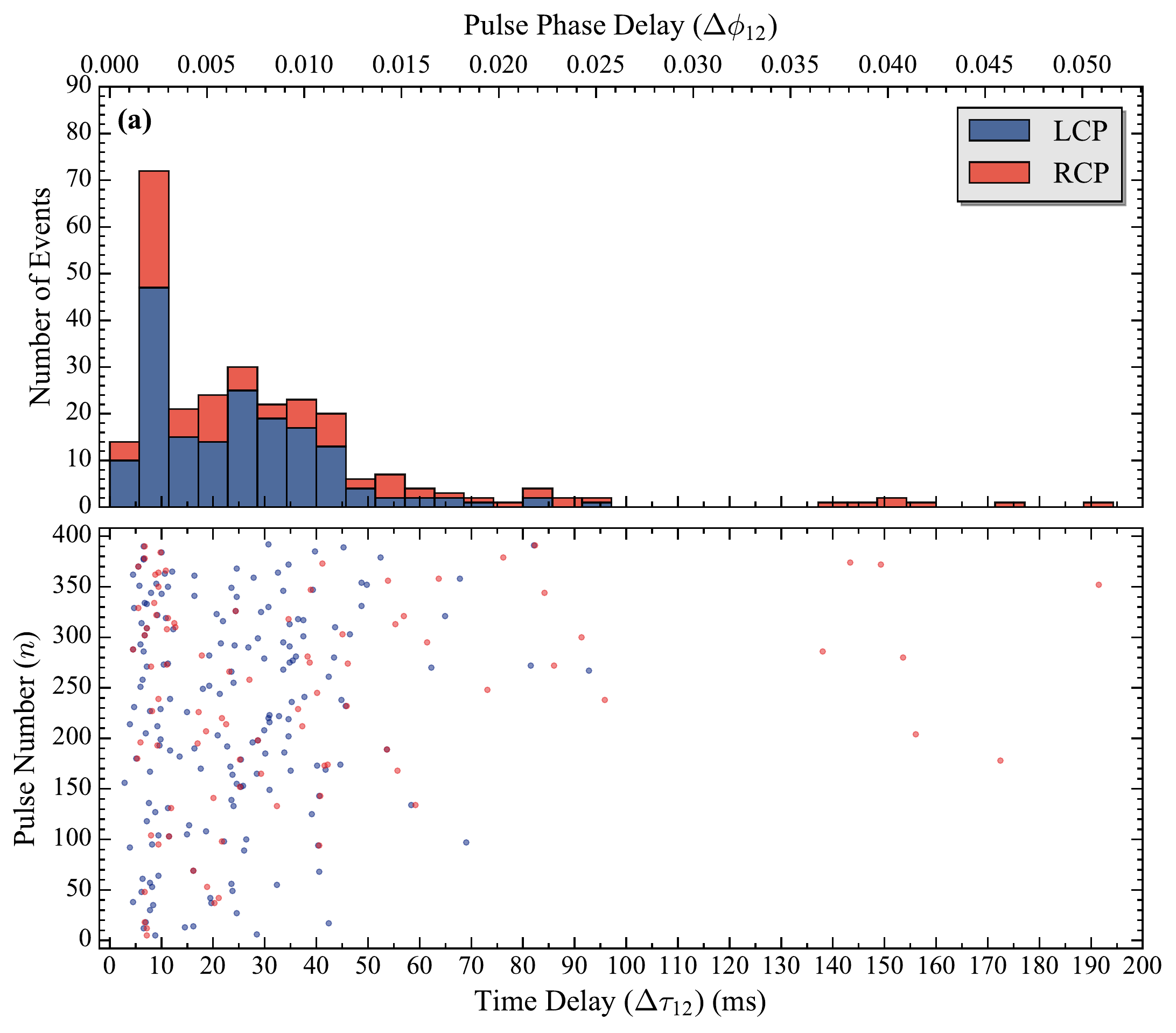}
			\label{Figure:Figure9a}
		}
		
		&
		
		\subfigure
		{
			\includegraphics[trim=0cm 0cm 0cm 0cm, clip=false, scale=0.39, angle=0]{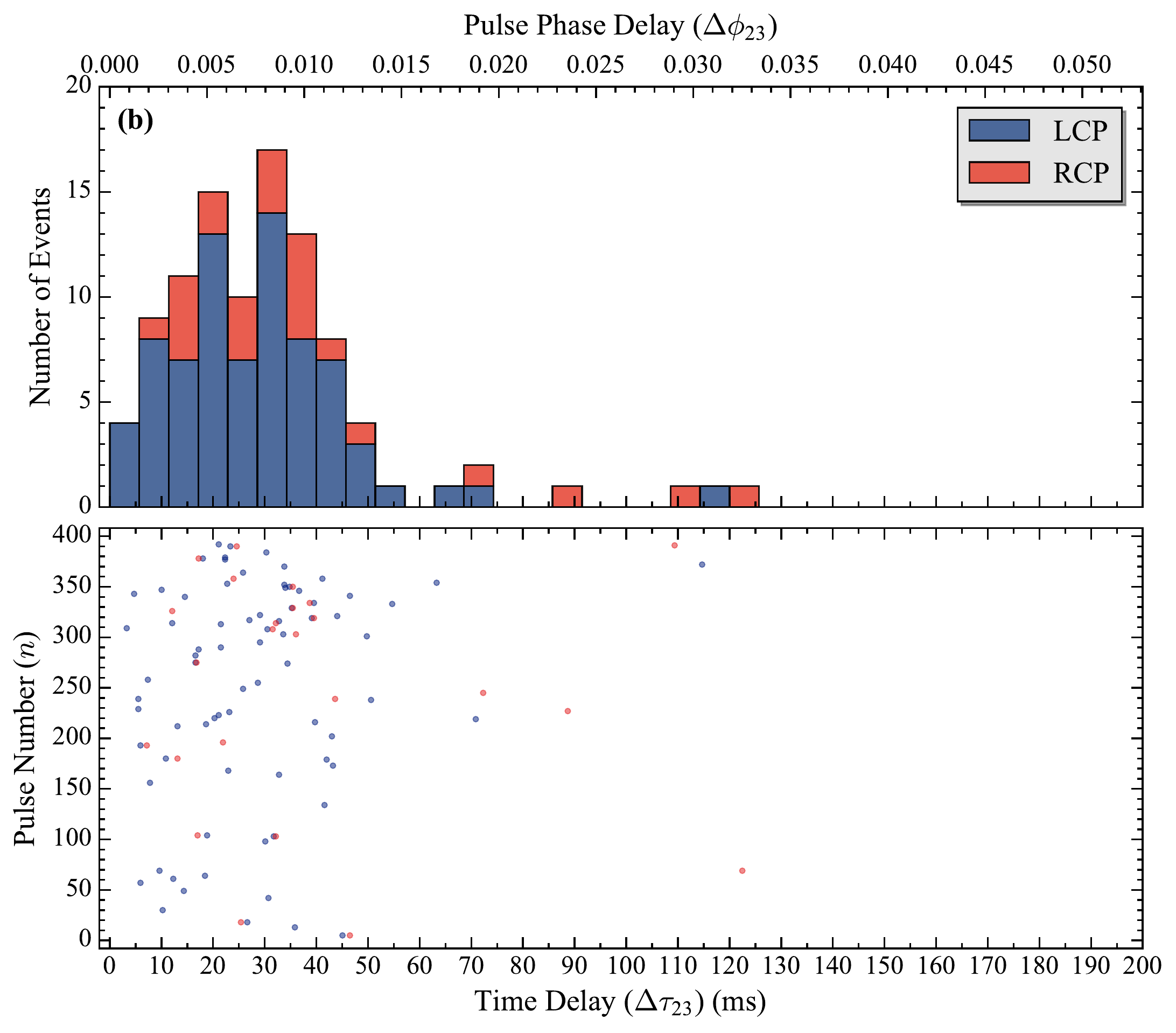}
			\label{Figure:Figure9b}
		}
		
	\end{tabular}
	
	\caption{Time delays between the \mbox{$X$-band} single pulse emission components detected in the (blue)~left circular polarization~(LCP) and (red)~right circular polarization~(RCP) channels during epoch~3. The emission component detected earliest in pulse phase during a given pulsar rotation is denoted by ``1" and emission components with later times of arrival~(ToAs) during the same pulse cycle are labeled sequentially. We show the time delays between emission components (a)~``1" and ``2" and (b)~``2" and ``3." Histograms of the time delays between the emission components are shown in the top panels, and the bottom panels show the distribution of the time delays measured from each polarization channel. Pulse numbers are referenced with respect to the start of the observation.}
	\label{Figure:Figure9}
\end{figure*}

We calculated characteristic time delays between the single pulse emission components by measuring time differences between adjacent components. We denote the emission component detected earliest in pulse phase during a given pulse cycle by ``1" and sequentially label subsequent emission components during the same pulse cycle. The distribution of time delays between the first two emission components was bimodal~(see~Figure~\ref{Figure:Figure9}(a)), which we associate with two distinct populations of single pulses. We report characteristic time delays of $\langle\tau_{\text{12}}\rangle_{\alpha}$\,$=$\,7.7\,ms and $\langle\tau_{\text{12}}\rangle_{\beta}$\,$=$\,39.5\,ms between the first two emission components from the mean delay of these separate distributions. Additionally, we measured a characteristic time delay of $\langle\tau_{\text{23}}\rangle$\,$=$\,30.9\,ms from the distribution of time delays between the second two components in Figure~\ref{Figure:Figure9}(b). We find that $\langle\tau_{\text{12}}\rangle_{\alpha}$ is comparable to the $\sim$10\,ms separation between components in the single pulses detected by the phased~VLA at 8.7\,GHz~\citep{Bower2014}.



\paragraph{Pulse Broadening}
\label{Section:Pulse_Broadening}

\begin{figure*}[t]
	\centering
	\includegraphics[trim=0cm 0cm 0cm 0cm, clip=false, scale=0.365, angle=0]{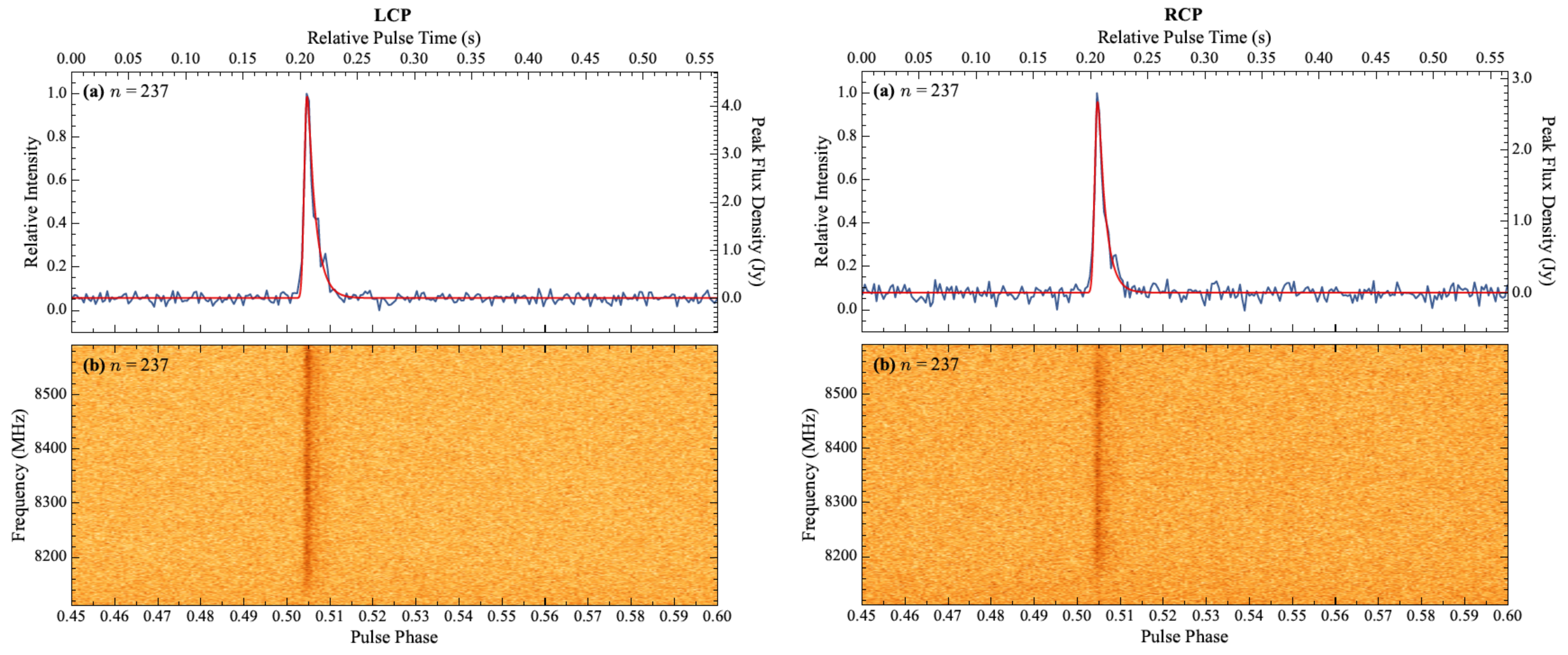}
	\caption{Bright \mbox{$X$-band} single pulse event displaying significant pulse broadening during pulse cycle~$n$\,$=$\,237 of epoch~3. The left and right plots show the detection of the single pulse in the left circular polarization~(LCP) and right circular polarization~(RCP) channels, respectively. We show the (a)~integrated single pulse profiles and (b)~dynamic spectra dedispersed using a DM~of~1778\,pc\,cm$^{\text{--3}}$ from both polarizations with a time resolution of 2\,ms. The \text{best-fit} scattering model, obtained by individually fitting the~LCP and RCP~data with Equation~(\ref{Equation:ObservedPulseShape}), is overlaid in red. The pulse broadening timescales measured for this event in each of these two polarization channels are $\tau_{d}^{\text{LCP}}$\,=\,7.1\,$\pm$\,0.2\,ms and $\tau_{d}^{\text{RCP}}$\,=\,6.7\,$\pm$\,0.3\,ms, respectively.}
	\label{Figure:Figure10}
\end{figure*}

The \mbox{$X$-band} single pulses detected during epochs~\text{1--4} displayed features characteristic of pulse broadening. In particular, many of the single pulse events showed significant evidence of exponential tails in their emission components (e.g.,~Figure~\ref{Figure:Figure10}), which is typically attributed to multipath scattering through the ISM~\citep{Williamson1972}. Strong exponential tails were sometimes observed in only a subset of the emission components during a given single pulse, with no pulse broadening in the other components (e.g.,~bottom row of Figure~\ref{Figure:Figure12}). A reverse exponential tail structure was also seen in some single pulse emission components (e.g.,~top and bottom rows of Figure~\ref{Figure:Figure12}), which may be due to more exotic pulse broadening mechanisms. The observed pulse broadening behavior could be explained by scattering from plasma inhomogenities, either local to or distant from the magnetar, or by unresolved low amplitude emission components.

Here, we measure a characteristic single pulse broadening timescale, $\tau_{d}$, at \mbox{$X$-band}~(8.4\,GHz) using the bright single pulse event shown in Figure~\ref{Figure:Figure10} from epoch~3. The amount of pulse broadening observed in this pulse is representative of other pulses with strong exponential tails. We use a thin scattering screen model, described in detail in the~\nameref{Section:Appendix}, to characterize the intrinsic properties of the pulse and the pulse broadening magnitude.

We fit the single pulse profiles in both polarization channels, shown in Figure~\ref{Figure:Figure10}(a), with Equation~(\ref{Equation:ObservedPulseShape}). A Bayesian Markov~chain Monte~Carlo~(MCMC) procedure was used to perform the fitting and incorporate covariances between the model parameters into their uncertainties. We assumed uninformed, flat priors on all of our model parameters and used a Gaussian likelihood function, $\mathcal{L}\propto\text{exp}(-\chi^{2}/2)$, such that the \text{log-likelihood} is given by:
\small
\begin{equation}
\ln\mathcal{L}=-\frac{1}{2}\left[\sum_{i}^{N}\left(\frac{P_{\text{obs},\,i}-P_{\text{model},\,i}}{\sigma_{i}}\right)^{2}+\ln(2\pi\sigma_{i}^{2})\right],
\label{Equation:GaussianLikelihood}
\end{equation}
\normalsize
where $N$ is the number of time bins in the single pulse profile, and $P_{\text{obs},i}$ and $P_{\text{model},i}$ are the measured and predicted values of the single pulse profile at bin~$i$, respectively. Each data point in the fit was weighted by the \text{off-pulse} RMS~noise level,~$\sigma_{i}$.

The posterior~PDFs of the model parameters in Equation~(\ref{Equation:ObservedPulseShape}) were sampled using an affine-invariant MCMC ensemble sampler~\citep{Goodman2010}, implemented in \texttt{emcee}\footnote{See https://dfm.io/emcee/current.}~\citep{ForemanMackey2013}. The parameter spaces were explored using 200~walkers and a chain length of 10,500 steps per walker. The first 500~steps in each chain were treated as the initial burn-in phase and were removed. \text{Best-fit} values for the model parameters were determined from the median of the marginalized posterior distributions using the remaining 10,000~steps in each chain, and uncertainties on the model parameters were derived from 1$\sigma$ Bayesian credible intervals.

\begin{figure*}[t]
	\centering
	\begin{tabular}{ccc}
		
		\subfigure
		{
			\includegraphics[trim=0cm 0cm 0cm 0cm, clip=false, scale=2.45, angle=0]{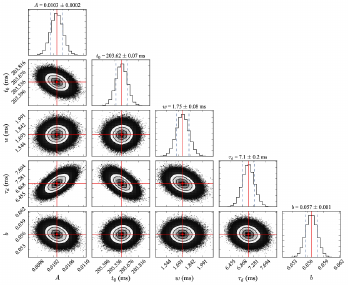}
			\label{Figure:Figure11a}
		}
		
		&
		
		\subfigure
		{
			\includegraphics[trim=0cm 0cm 0cm 0cm, clip=false, scale=2.45, angle=0]{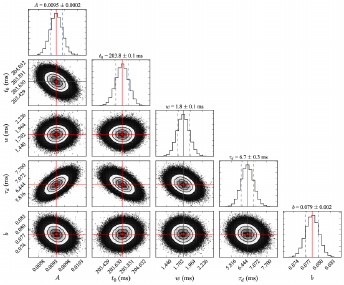}
			\label{Figure:Figure11b}
		}
		
	\end{tabular}
	
	\caption{Corner plots showing the marginalized posterior distributions obtained by independently fitting the integrated single pulse profiles from the left circular polarization~(LCP) and right circular polarization~(RCP) channels in Figure~\ref{Figure:Figure10}(a) with the scattering model in Equation~(\ref{Equation:ObservedPulseShape}). The posterior distributions shown on the left and right are derived from fitting the~LCP and RCP~data, respectively. Single parameter projections of the posterior probability distributions and \text{best-fit} values are shown along the diagonal, and the off-diagonal plots are the marginalized two-dimensional posterior distributions. Covariances between the model parameters are indicated by a tilted error ellipse. The red lines correspond to the \text{best-fit} values for the model parameters derived from the median of the single parameter posterior distributions, and the dashed blue lines indicate 1$\sigma$ Bayesian credible intervals.}
	\label{Figure:Figure11}
\end{figure*}

\begin{figure*}[t]
	\centering
	\begin{tabular}{ccc}
		
		\subfigure
		{
			\includegraphics[trim=0cm 0cm 0cm 0cm, clip=false, scale=0.36, angle=0]{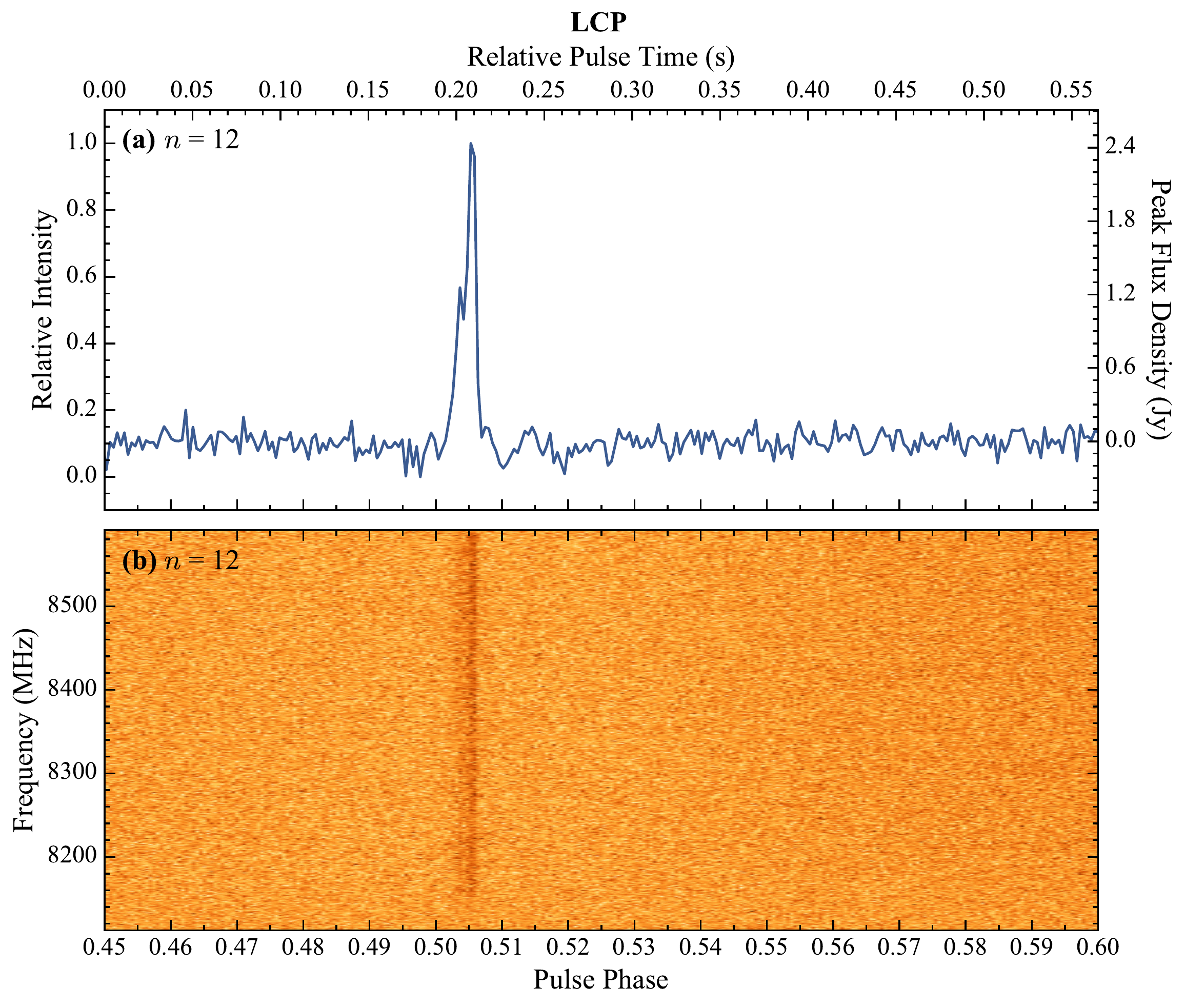}
			\label{Figure:Figure12a}
		}
		
		&
		
		\subfigure
		{
			\includegraphics[trim=0cm 0cm 0cm 0cm, clip=false, scale=0.36, angle=0]{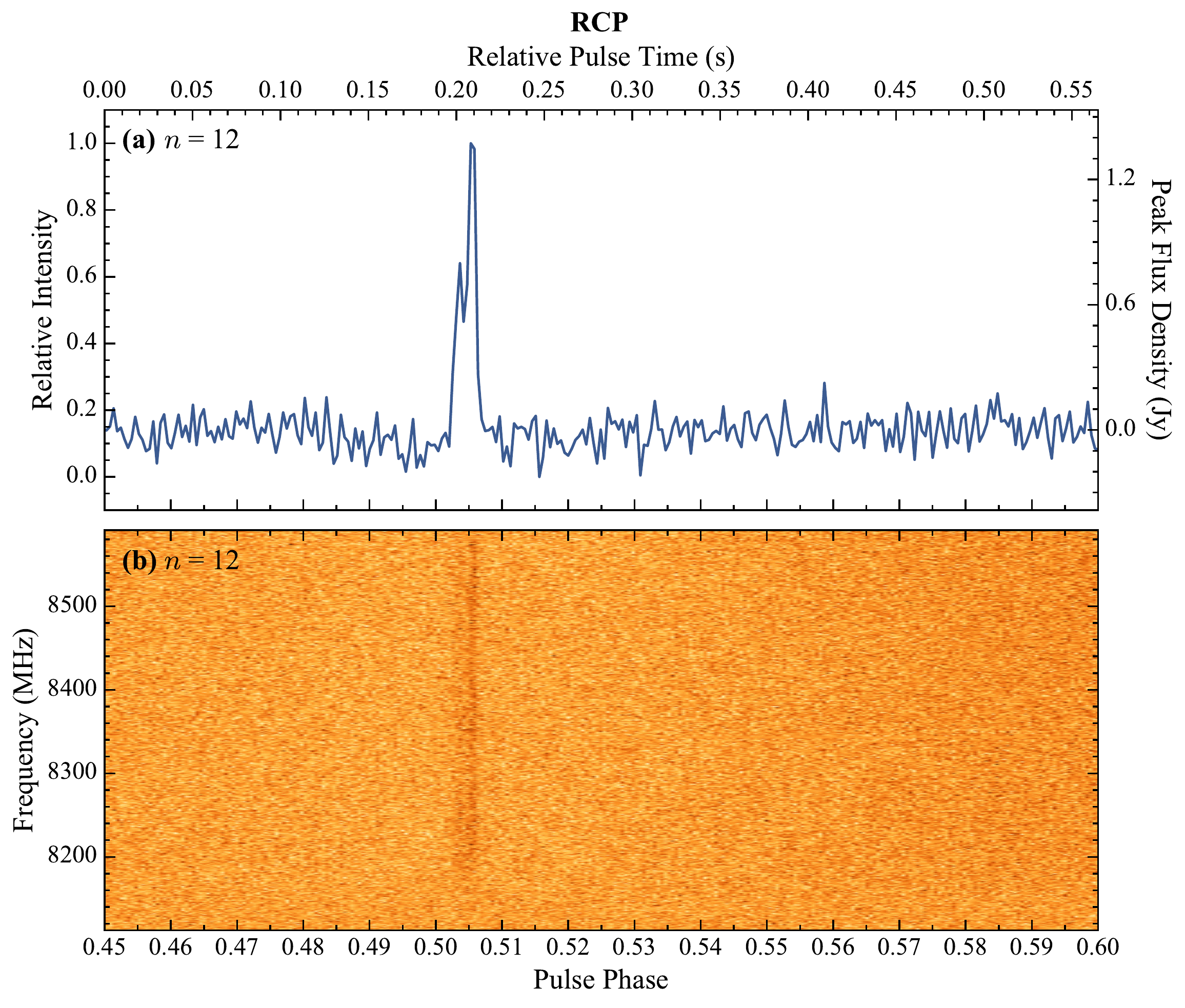}
			\label{Figure:Figure12b}
		}
		
		\\
		
		\subfigure
		{
			\includegraphics[trim=0cm 0cm 0cm 0cm, clip=false, scale=0.36, angle=0]{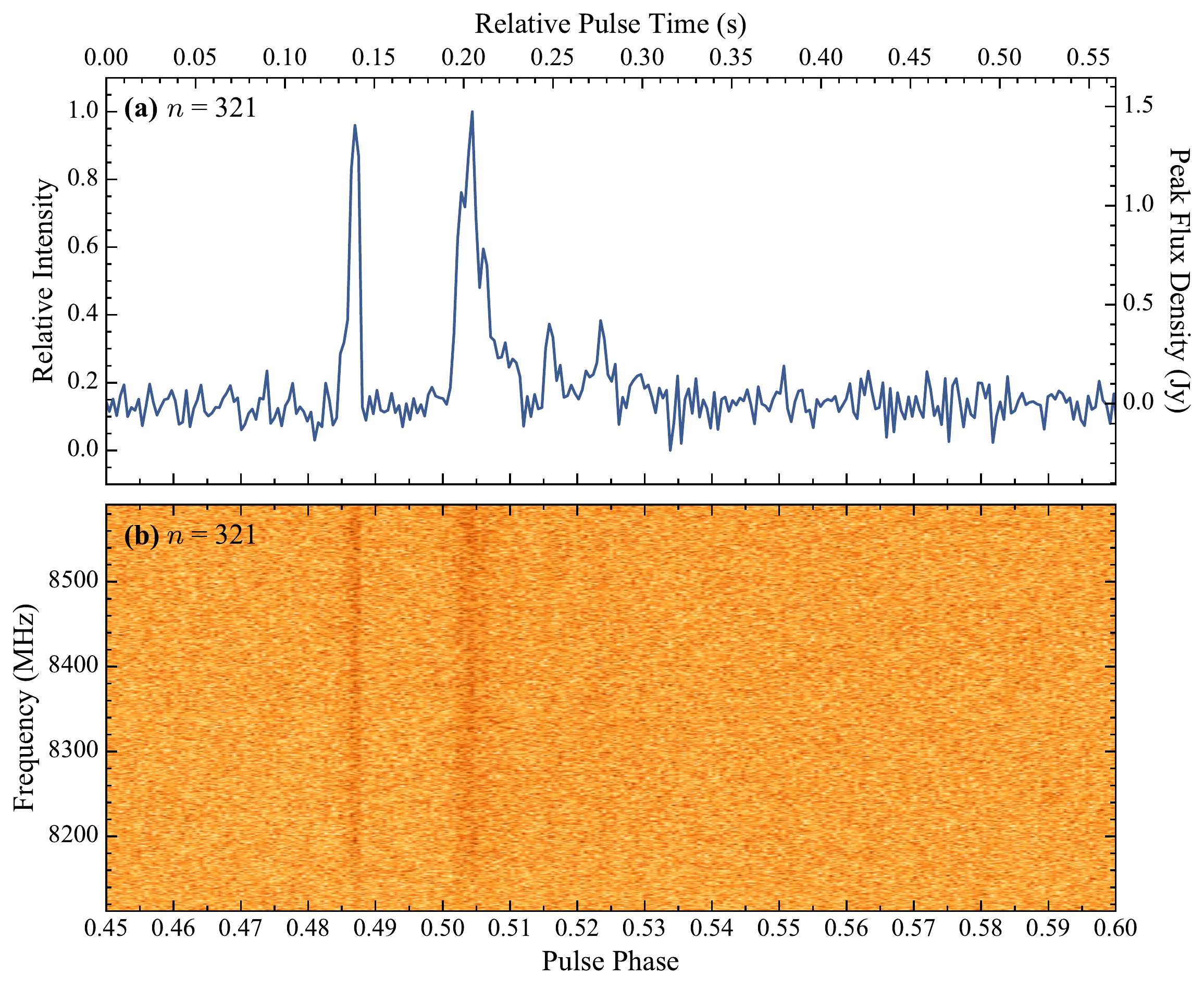}
			\label{Figure:Figure12c}
		}
		
		&
		
		\subfigure
		{
			\includegraphics[trim=0cm 0cm 0cm 0cm, clip=false, scale=0.36, angle=0]{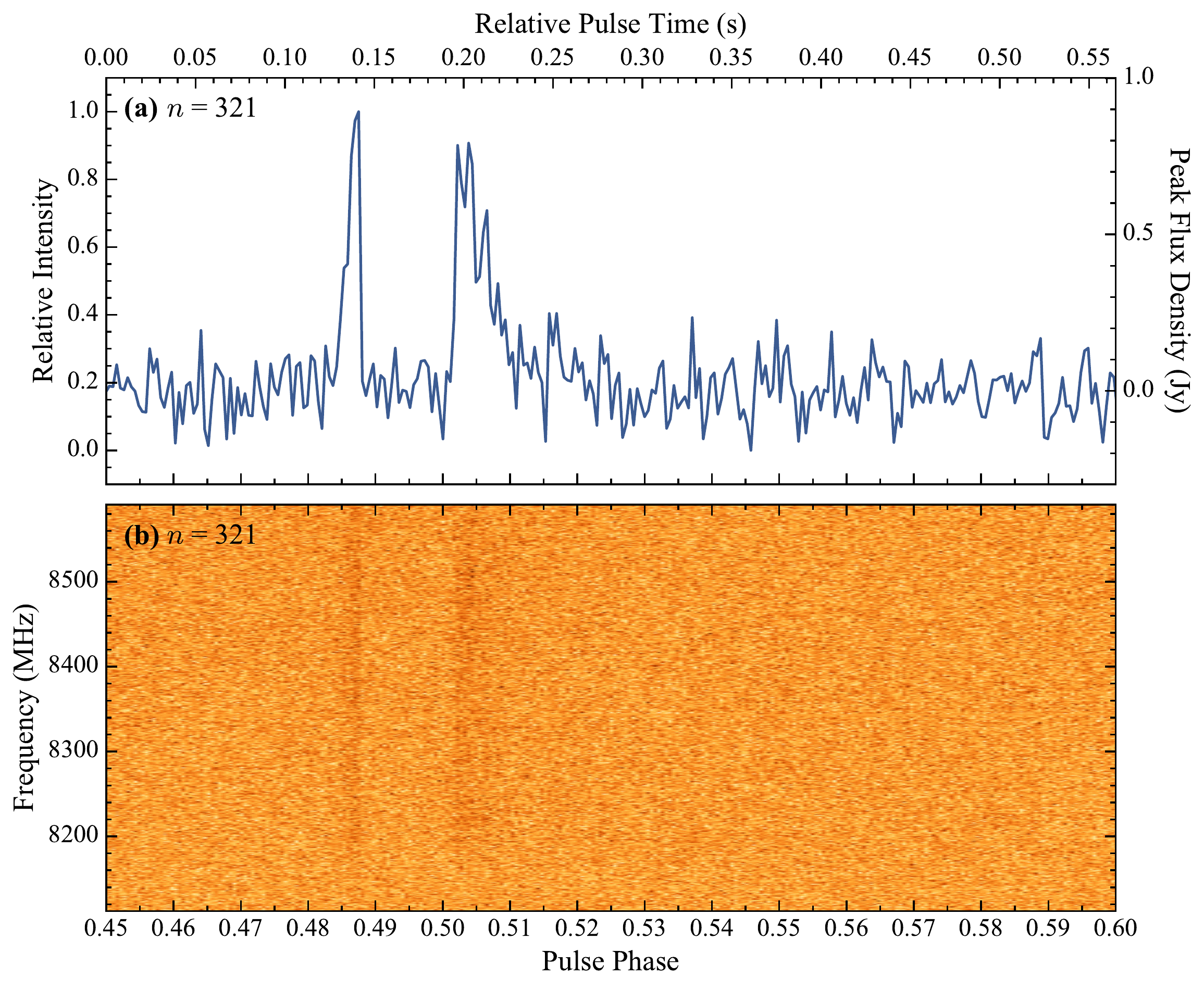}
			\label{Figure:Figure12d}
		}
		
	\end{tabular}
	
	\caption{Examples of bright \mbox{$X$-band} single pulse events displaying exotic pulse broadening behavior during pulse cycles (top~row)~$n$\,$=$\,12 and (bottom~row)~$n$\,$=$\,321 of epoch~3. The plots in the left and right columns show detections of the single pulses in the left circular polarization~(LCP) and right circular polarization~(RCP) channels, respectively. We show the (a)~integrated single pulse profiles and (b)~dynamic spectra dedispersed using a DM~of~1778\,pc\,cm$^{\text{--3}}$ from both polarizations with a time resolution of 2\,ms. The pulse shapes of the dominant emission components from pulse cycles~$n$\,$=$\,12 and~$n$\,$=$\,321 resemble a reverse exponential tail. The secondary emission component detected during pulse cycle~$n$\,$=$\,321 has a traditional scattering tail shape, which is not observed in the other emission components.}
	\label{Figure:Figure12}
\end{figure*}

The \text{best-fit} scattering model is overlaid in red on the single pulse profiles in Figure~\ref{Figure:Figure10}(a) for each polarization channel, and we show the marginalized posterior distributions in the corner plots in Figure~\ref{Figure:Figure11} from individually fitting the~LCP and RCP~data. Pulse broadening timescales of $\tau_{d}^{\text{LCP}}$\,=\,7.1\,$\pm$\,0.2\,ms and $\tau_{d}^{\text{RCP}}$\,=\,6.7\,$\pm$\,0.3\,ms were obtained for this single pulse event from the single parameter marginalized posterior distributions, and they are consistent with each other to within 1$\sigma$. We report a characteristic pulse broadening timescale of $\langle\tau_{d}\rangle$\,=\,6.9\,$\pm$\,0.2\,ms from averaging these two independent polarization channel measurements. These values are comparable to the characteristic time delay $\langle\tau_{\text{12}}\rangle_{\alpha}$ of~7.7\,ms between the leading two single pulse emission components reported in Section~\ref{Section:Temporal_Variability}, which may indicate that the exponential tails observed in the single pulses could be formed from multiple adjacent emission components.

Pulse broadening measurements by~\citet{Spitler2014} between 1.19~and~18.95\,GHz yielded a spectral index of \mbox{$\alpha_{d}$\,$=$\,--3.8\,$\pm$\,0.2} and a scattering timescale of $\tau_{d}$\,$=$\,1.3\,$\pm$\,0.2\,s at 1\,GHz, which implies $\tau_{d}$\,$\approx$\,0.4\,ms at 8.4\,GHz. If the exponential tail structure in the single pulse events were produced by scattering through a thin screen in the~ISM, then our characteristic pulse broadening timescale of 6.9\,ms suggests that individual single pulse events can have scattering timescales that are more than an order of magnitude larger than the scattering predicted at this frequency by~\citet{Spitler2014}. We also point out that our pulse broadening measurements are significantly larger than the scattering timescale predicted at this frequency by~\citet{Bhat2004}, where a mean spectral index of \mbox{$\alpha_{d}$\,=\,--3.9\,$\pm$\,0.2} was derived from integrated pulse profiles between 0.43~and~2.38\,GHz of low Galactic latitude pulsars with moderate~DMs. An earlier study of nine highly dispersed pulsars between 0.6~and~4.9\,GHz yielded a spectral index of \mbox{$\alpha_{d}$\,=\,--3.44\,$\pm$\,0.13}~\citep{Lohmer2001}, but this is also too steep to account for the amount of single pulse broadening seen here at \mbox{$X$-band}. Additionally, the level of pulse broadening reported here is inconsistent with a pure Kolmogorov spectrum, which has an expected spectral index of $\alpha_{d}$\,=\,--4.4~\citep{Lee1975}.



\section{Discussion and Conclusions}
\label{Section:Discussion}

\subsection{Intrinsic and Extrinsic Emission Characteristics}
\label{Section:Intrinsic_Extrinsic_Emission}

There are various timescales observed from the data that describe the emission:~(1)~the typical intrinsic width of individual emission components~($w$\,$=$\,1.8\,ms), (2)~a characteristic pulse broadening timescale~($\langle\tau_{d}\rangle$\,=\,6.9\,ms), (3)~a prevailing delay time between successive components~($\langle\tau_{\text{12}}\rangle_{\alpha}$\,$=$\,7.7\,ms), (4)~the envelope of pulse delays between successive components~($\Delta_{\text{12}}$\,$\approx$\,50\,ms), (5)~the spread of component arrival times relative to the magnetar rotation period~($\Delta_{\text{comp}}$\,$\approx$\,100\,ms), and (6)~the magnetar rotation timescale~($P$\,$=$\,3.77\,s).

The single pulse morphology may be caused by processes that are either intrinsic or extrinsic to the magnetar. The variability in pulse structure between emission components argues for an intrinsic origin, though external scattering or refractive lensing could also be responsible. Extrinsic mechanisms would have to produce fast line of sight changes on millisecond to second timescales, which suggests that such structures are located near the~magnetar.

Pulses were observed during more than 70\% of the magnetar's rotations, but often not at precisely the same phase (see Figures~\ref{Figure:Figure7}--\ref{Figure:Figure9}). This implies that the active site of the emission must emit pulses fairly continuously since pulses were seen during almost all rotations. The 3--50\,ms timescale between successive components is likely indicative of the pulse repetition rate. The width of the distribution of pulse components is approximately $\pm$0.02 in phase units~(see Figure~\ref{Figure:Figure7}). These observations are most naturally explained by fan beam emission with a width of about $\pm$7$^{\circ}$. The data are consistent with a single primary active region of emission since pulse components were generally not detected outside of a narrow phase range, except in a few cases.

Figure~\ref{Figure:Figure4} shows that most rotations exhibited multiple pulse components, although single components were not uncommon. Giant pulses were detected primarily during the second half of epoch~3, which indicates that these bright events are transient in nature. There is some evidence that the brightest pulse component appears first during a given rotation, occasionally followed by weaker components. This may indicate that the active region can have outbursts that trigger additional bursts. Alternatively, the dimming of later pulse components may be due to effects from tapering of the fan beam.

Recent measurements of \text{PSR~J1745--2900's} linear polarization fraction showed large variations as a function of time (see~Figure~3 in~\citealt{Desvignes2018}). We note that the strength of the single pulse emission during epochs~\text{1--4} seems to roughly coincide with changes in the polarization fraction. In particular, the third epoch showed the strongest emission components during one of the periods of maximum linear polarization. \citet{Desvignes2018} discuss whether the frequency dependent polarization behavior could be intrinsic or extrinsic to the magnetar.

\subsection{Scattering Regions}
\label{Section:Scattering_Regions}

Hyperstrong radio wave scattering from pulsars near the~GC has typically been modeled by a single thin scattering screen~\citep{Cordes1997, Lazio1998}. The amount of pulse broadening produced by multipath propagation through the screen depends on its distance from the~GC~($\Delta_{\text{GC}}$), which can be calculated from~\citep{Cordes1997}:
\begin{equation}
\Delta_{\text{GC}}=\frac{D_{\text{GC}}}{1+\left(\frac{\tau_{d}}{6.3\,\text{s}}\right)\left(\frac{8.5\,\text{kpc}}{D_{\text{GC}}}\right)\left(\frac{1.3\,\text{arcsec}}{\theta_{1\,\text{GHz}}}\right)^{2}\left(\frac{\nu}{1\,\text{GHz}}\right)^{4}},
\label{Equation:ScatteringScreen}
\end{equation}
where $\tau_{d}$ is the temporal scattering timescale at an observing frequency $\nu$, $D_{\text{GC}}$\,$=$\,8.3\,$\pm$\,0.3\,kpc is the distance to the~GC~\citep{Gillessen2009}, and $\theta_{\text{1\,GHz}}$\,$=$\,1075\,$\pm$\,50\,mas is the angular size of \text{PSR~J1745--2900} scaled to 1\,GHz~\citep{Bower2014}. Assuming the pulse broadening reported in Section~\ref{Section:Pulse_Broadening} is entirely due to temporal scattering from diffraction through a single thin screen, Equation~(\ref{Equation:ScatteringScreen}) can be used to determine the screen's distance from the~GC. We find that a scatter broadening timescale of $\langle\tau_{d}\rangle$\,$=$\,6.9\,$\pm$\,0.2\,ms at 8.4\,GHz would require a screen at a distance of 0.9\,$\pm$\,0.1\,kpc from the~GC.

The number of scattering screens and their locations is an important consideration, which has strong implications on searches for pulsars toward the~GC. A screen is thought to exist at a distance of $\sim$5.8\,kpc from the~GC based on temporal broadening measurements between 1.2~and~8.7\,GHz~\citep{Bower2014}. \citet{Wucknitz2014} argued that most of the temporal and angular scattering from this magnetar are produced by a single thin scattering screen $\sim$4.2\,kpc from the pulsar. However, a single screen at either of these distances is incompatible with the single pulse broadening reported here at 8.4\,GHz if the broadening is attributed to thin screen scattering. A distant, static scattering screen also cannot account for the variations in broadening between components on short timescales. Scattering from regions much closer to the~GC~($<$\,1\,kpc) have also been proposed~\citep{Lazio1998, Dexter2017}, but a single \text{sub-kiloparsec} screen overestimates the amount of scattering reported by~\citet{Spitler2014} between 1.19~and~18.95\,GHz.

Recently, \citet{Desvignes2018} observed rapid changes in the magnetar's~RM and depolarized radio emission at 2.5\,GHz. They attributed the variations in RM to \text{magneto-ionic} fluctuations in the GC and explained the depolarization behavior by invoking a secondary scattering screen at a distance of $\sim$0.1\,pc in front of the magnetar, assuming a screen size of $\sim$1.9\,au and a scattering delay time of $\sim$40\,ms. A two scattering screen model, consisting of a local screen~($<$\,700\,pc from the~GC) and a distant screen~($\sim$\,5\,kpc from the~GC), has also been proposed to explain the angular and temporal broadening from other GC~pulsars, which cannot be accounted for by a single scattering medium~\citep{Dexter2017}. In the case of \text{PSR~J1745--2900}, \citet{Dexter2017} argued that a local screen would not significantly contribute to the temporal broadening. On the other hand, a two component scattering screen, with a strong scattering central region and weak scattering extended region, may explain both the $\tau_{d}$\,$\propto$\,$\nu^{\text{--3.8}}$ temporal scattering at lower frequencies~\citep{Spitler2014} and larger broadening times at higher frequencies. Depending on the scattering strengths and sizes of the regions, the spectrum of pulse broadening times can flatten at higher frequencies (e.g.,~see~Figure~3 in~\citealt{Cordes2001}).

Strong variability in the single pulse broadening times was seen on short timescales between pulse cycles and individual emission components within the same pulse cycle~(see~Figures~\ref{Figure:Figure3},~\ref{Figure:Figure10}, and~\ref{Figure:Figure12}). \text{Ultra-fast} changes in the scattering media on roughly millisecond to second timescales would be required to explain this variability using multiple screens. Scattering regions formed from turbulent, \text{fast-moving} plasma clouds in close proximity to the magnetar might be one possible mechanism that could produce this variability. Similar models have been used to explain the temporal structure of pulses from the Crab pulsar~\citep{Lyne1975, Crossley2010}. Alternatively, this behavior could be explained by an ensemble of plasma filaments near a strong scattering screen close to the magnetar, where these filaments create inhomogeneities in the scattering medium~\citep{Cordes2001}.

We also consider the possibility that the single pulse broadening is intrinsic in origin. Pulsed radio emission from magnetars is known to be highly variable, and strong spiky single pulses have been observed from other radio magnetars~(e.g.,~\text{PSR~J1622--4950};~\citealt{Levin2012}). The similarity between the time delay between single pulse emission components, $\langle\tau_{\text{12}}\rangle_{\alpha}$\,$\approx$\,8\,ms, and the single pulse broadening timescale, $\langle\tau_{d}\rangle$\,$\approx$\,7\,ms, suggests that exponential tails in the single pulses may be comprised of multiple unresolved adjacent components.

\subsection{Plasma Clouds and Plasma Lenses}
\label{Section:Plasma_Clouds_Plasma_Lenses}

First, we describe how multipath propagation through compact, high density plasma clouds may give rise to variable pulse broadening between pulse components. This behavior could be produced during pulse cycles where one or more of these clouds traverse the radio beam at high velocities. Inhomogeneities in the clouds could result in different observed scattering shapes. These objects would also have to be transient to explain the differences in broadening between components in the same pulse cycle, which argues for locations near the pulsar magnetosphere. These plasma clouds are postulated to exist in the physical environment of the magnetosphere.

We provide estimates of the temperature~($T_{\text{PC}}$) of the plasma cloud~(PC), smallest elementary thickness of the inter-plasmoid current layer~($\delta$), and distance from the magnetar~($D_{\text{PC}}$) where these structures are expected to exist. Our calculations follow the model in~\citet{Uzdensky2014}, which assumes magnetic reconnection occurs in the pulsar magnetosphere and allows for strong optically thin synchrotron radiative cooling inside the layer. We assume a canonical neutron star mass of $M_{\star}$\,$=$\,1.4\,$M_{\odot}$, with radius $R_{\star}$\,$=$\,10\,km and moment of inertia $I$\,$=$\,10$^{\text{45}}$\,g\,cm$^{\text{2}}$. The magnetar's surface dipolar magnetic field is $B_{\text{surf}}$\,$\approx$\,3.2\,$\times$\,10$^{\text{19}}$$(P\dot{P})^{1/2}$\,G\,$\approx$\,2.6\,$\times$\,10$^{\text{14}}$\,G~\citep{Lynch2015}, which is $\sim$6 times larger than the quantum critical magnetic field, $B_{\text{Q}}$\,$=$\,$m_{e}^{2}c^{3}/e\hbar$\,$\approx$\,4.4\,$\times$\,10$^{\text{13}}$\,G. If we assume the magnetic field is approximately dipolar inside the light cylinder~(LC), we can estimate the magnetic field at distances $D_{\text{PC}}$\,$\le$\,$R_{\text{LC}}$\,$=$\,$cP/2\pi$\,$\approx$\,1.8\,$\times$\,10$^{\text{5}}$\,km using $B_{\text{PC}}$\,$\approx$\,$B_{\text{surf}}(R_{\star}/R_{\text{PC}})^{3}$. Magnetic reconnection likely occurs at distances much smaller than the light cylinder radius since the predicted magnetic field is considerably weaker at the edge of the light cylinder~($B_{\text{LC}}$\,$\approx$\,2.9\,$\times$\,10$^{\text{8}}P^{-5/2}\dot{P}^{1/2}$\,G\,$\approx$\,45\,G). Following the analysis in~\citet{Uzdensky2014}, we find that, at a distance of $D_{\text{PC}}$\,$=$\,5000\,km\,$=$\,500$R_{\star}$\,$\sim$\,0.03$R_{\text{LC}}$, the magnetic field inside the magnetosphere is $B_{\text{PC}}$\,$\approx$\,2\,$\times$\,10$^{\text{6}}$\,G. If this field is comparable to the reconnecting magnetic field in the comoving frame of the relativistic pulsar wind, then plasma clouds formed in the pulsar's magnetosphere can have densities of $n_{\text{PC}}$\,$\sim$\,10$^{\text{12}}$\,cm$^{\text{--3}}$, with temperatures of $T_{\text{PC}}$\,$\sim$\,50\,GeV, and plasma scales of order $\delta$\,$\sim$\,150\,cm or larger (in the comoving frame). 

Scattering from plasma clouds may give rise to pulse broadening by an amount $\tau_{d}$\,$\sim$\,$L_{\text{PC}}^{\text{2}}$/2$c$$D_{\text{PC}}$, where $L_{\text{PC}}$ is the cloud size. In order to explain a broadening timescale of $\sim$7\,ms, such clouds can be no larger than $L_{\text{PC}}$\,$\sim$\,4600\,km, assuming a distance of $D_{\text{PC}}$\,$=$\,5000\,km to the cloud. Although different cloud geometries are possible, these estimates suggest that high density plasma clouds in close proximity to the pulsar wind could produce changes on the short timescales needed to explain the pulse broadening variations between emission components.

Next, we discuss possible mechanisms responsible for the frequency structure observed in the single pulse emission components. During most magnetar rotations, individual pulse components showed variations in brightness with frequency, but all of the components were not strongly affected simultaneously during pulse cycles where the emission was multipeaked. In many cases, the pulsed radio emission vanished over a significant fraction of the frequency bandwidth~(e.g.,~Figures~\ref{Figure:Figure3}~and~\ref{Figure:Figure5}). We argue that these effects are likely extrinsic to the magnetar and can be produced by strong lensing from refractive plasma structures, but may also be intrinsic to the magnetar.

Lensing from structures near the magnetar may account for the variations in brightness between closely spaced components. This mechanism has also been proposed to explain echoes of radio pulses from the Crab pulsar~\citep{Backer2000, GrahamSmith2011} and bursts from FRB~sources~\citep{Cordes2017}. Applying the model in~\citet{Cordes2017}, we find that a \text{one-dimensional} Gaussian plasma lens at a distance of $d_{\text{sl}}$\,$=$\,$R_{\text{LC}}$\,$=$\,1.8\,$\times$\,10$^{\text{5}}$\,km from the magnetar, with a scale size of $a$\,$\sim$\,5300\,km and lens dispersion measure depth of DM$_{\text{$\ell$}}$\,$\sim$\,10\,pc\,cm$^{\text{--3}}$, can produce frequency structure on scales of $\sim$1--500\,MHz near a focal frequency of $\sim$8.4\,GHz. Caustics can induce strong magnifications, with changes in gain spanning \text{1--2} orders of magnitude for this particular lens configuration. Frequency dependent interference effects are most prominent near the focal frequency and become attenuated at higher frequencies. Larger dispersion depths would result in higher focal frequencies, which is certainly a possibility for plasma lenses located near the~GC. Multiple plasma lenses may be responsible for the observed behavior, and they may have a variety of sizes, dispersion depths, and distances from the source that could differ from the parameter values considered here. Alternatively, this behavior could be intrinsic, possibly similar in nature to the banded structures observed in one of the components of the Crab pulsar, namely the \text{High-Frequency} Interpulse~\citep{Hankins2016}.

\subsection{Comparison with Other Magnetars and~High~Magnetic~Field~Pulsars}
\label{Section:Comparison_Magnetars_Pulsars}

\text{PSR~J1745--2900} shares remarkable similarities with the three other radio magnetars: \text{XTE~J1810--197}, \text{1E~1547.0--5408}, and \text{PSR~J1622--4950}~\citep{Camilo2006, Camilo2007a, Levin2010}. They all exhibit extreme variability in their pulse profiles, radio flux densities, and spectral indices, which are quite anomalous compared to ordinary radio pulsars. Their average pulse shapes and flux densities can also change on short timescales of hours to days~(e.g.~\citealt{Camilo2008, Camilo2016, Levin2010, Pennucci2015}). Radio pulses from these magnetars are typically built up of multiple spiky subpulses with widths on the order of milliseconds~\citep{Serylak2009, Levin2012} and can be exceptionally bright, with peak flux densities exceeding 10\,Jy in the case of \text{XTE~J1810--197}~\citep{Camilo2006}. The flux densities of these events are unlike the giant pulses observed from the Crab pulsar~\citep{Cordes2004}, which have an energy flux distribution that follows a power law~\citep{Majid2011}.

These magnetars tend to have relatively flat or inverted radio spectra, while ordinary radio pulsars have much steeper spectra on average (mean spectral index $\langle\alpha\rangle$\,$=$\,--1.8\,$\pm$\,0.2;~\citealt{Maron2000}). This makes the detection of normal radio pulsars challenging at frequencies above a few gigahertz. To date, only seven ordinary pulsars have been detected at frequencies above 30\,GHz~\citep{Wielebinski1993, Kramer1997, Morris1997, Lohmer2008}. In contrast, two of these radio magnetars (\text{PSR~J1745--2900} and \text{XTE~J1810--197}) have been detected at record high frequencies (291~and~144\,GHz, respectively;~\citealt{Camilo2007c, Torne2017}). Daily changes in the spectral indices of these four radio magnetars have also been observed~(e.g.~\citealt{Lazaridis2008, Anderson2012}). Unusually steep radio spectra have been obtained from \text{PSR~J1745--2900} and \text{XTE~J1810--197}~\citep{Pennucci2015, Camilo2016}, with negative spectral indices comparable to those reported here in Table~\ref{Table:FluxDensitiesSpectralIndices}.

The pulsed radio emission from magnetars is often highly linearly polarized~\citep{Camilo2007b, Levin2012}, but large variations in polarization have been seen~\citep{Desvignes2018}. With the exception of \text{XTE~J1810--197}, the other radio magnetars have~RMs that fall in the top 1\% of all known pulsar~RMs, indicating that they inhabit extreme \text{magneto-ionic} environments. Radio emission from magnetars can also suddenly shut off, and quiescence periods can last for many hundreds of days~(e.g.,~\citealt{Camilo2016, Scholz2017}), but no such behavior has yet been reported for the GC~magnetar. However, \text{PSR~J1745--2900} was not detected during searches for compact radio sources in the~GC or in \text{X-ray} scans of the Galactic~plane prior to its discovery~(e.g.,~\citealt{Lazio2008, Baumgartner2013}), which suggests that it was quiescent before its initial \text{X-ray} outburst in 2013~April~\citep{Kennea2013}.

\text{PSR~B1931+24}, an ordinary isolated radio pulsar, has exhibited quasi-periodic deactivation and reactivation of its radio emission on timescales of 25--35\,days~\citep{Kramer2006}, but this is extremely atypical of radio pulsars. Normal \text{rotation-powered} pulsars with high magnetic fields, such as \text{PSR~J1119--6127}, have displayed mode changes over days to weeks following \text{magnetar-like} \text{X-ray} outbursts, which resemble the emission characteristics of radio magnetars~\citep{Majid2017, Dai2018}. This suggests an underlying connection between ordinary pulsars and magnetars. To our knowledge, our observations of frequency dependent variations in the individual pulses of \text{PSR~J1745--2900}~(see~Section~\ref{Section:Frequency_Structure}) are the first examples of such behavior from a radio magnetar.

\subsection{Similarities with Fast Radio Bursts}
\label{Section:Similarities_FRBs}

As pointed out in various papers~(e.g.,~\citealt{Pen2015, Metzger2017, Michilli2018}), an extragalactic magnetar near a massive black hole could be the progenitor of~FRBs. There are numerous similarities between the emission from the GC~magnetar and FRB~sources, such as the repeating FRB~121102. High dispersion measures are observed from both FRB~121102~(DM\,$=$\,560\,pc\,cm$^{\text{--3}}$;~\citealt{Michilli2018}) and \text{PSR~J1745--2900}~(DM\,$=$\,1778\,pc\,cm$^{\text{--3}}$). These objects also exhibit large, variable~RMs (RM\,$\approx$\,1.4\,$\times$\,10$^{\text{5}}$\,rad\,m$^{\text{--2}}$/(1\,$+$\,$z$)$^{\text{2}}$, where $z$\,$\sim$\,0.2 for FRB~121102;~\citealt{Michilli2018}, compared to \mbox{RM\,$\approx$\,--7\,$\times$\,10$^{\text{4}}$\,rad\,m$^{\text{--2}}$} for the GC~magnetar;~\citealt{Eatough2013b, Desvignes2018}). Multicomponent bursts with widths $\lesssim$\,1\,ms have been reported from FRB~121102~\citep{Gajjar2018, Michilli2018}. This is similar to the pulse morphology of the GC~magnetar, which shows emission components with comparable pulse widths that are likely broadened. In both cases, the detected burst spectra show frequency structure on similar scales, which may be produced by the same underlying mechanism~(e.g.,~\citealt{Spitler2016, Gajjar2018, Michilli2018}). Other FRBs, such as FRB~170827~\citep{Farah2018}, exhibit bursts with frequency structure on much finer scales ($\lesssim$\,2\,MHz) at frequencies below $\sim$1\,GHz.

At a luminosity distance of $\sim$1\,Gpc, the energy output of bursts from FRB~121102 is a factor of $\sim$10$^{\text{10}}$ larger than the single pulse emission from \text{PSR~J1745--2900}. However, we find that strong focusing by a single plasma lens can produce caustics that may boost the observed flux densities of bursts from FRB~121102 by factors of 10--10$^{\text{6}}$ on short timescales~\citep{Cordes2017}. Multiple plasma lenses could yield even larger burst magnifications. We also note that many pulses from the GC~magnetar had $\gtrsim$\,10 times the typical single pulse intensity, and the emission rate of these giant pulses was time-variable. Therefore, an~FRB source like FRB~121102 could possibly be an extreme version of a magnetar, such as \text{PSR~J1745--2900}.


\section*{Acknowledgments}

We thank the referee for valuable comments that helped us improve this paper. We also thank Professor~Roger~Blandford for insightful discussions and suggestions.

A.~B.~Pearlman acknowledges support by the Department of Defense~(DoD) through the National Defense Science and Engineering Graduate Fellowship~(NDSEG) Program and by the National Science Foundation Graduate Research Fellowship under Grant~No.~\text{DGE-1144469}.

We thank the Jet~Propulsion~Laboratory and Caltech's President's and Director's Fund for partial support at~JPL and the Caltech~campus.  We also thank Joseph Lazio and Charles Lawrence for providing programmatic support for this work. A portion of this research was performed at the Jet~Propulsion~Laboratory, California~Institute~of~Technology and the Caltech campus, under a Research and Technology Development Grant through a contract with the National Aeronautics and Space Administration. U.S.~government sponsorship is acknowledged.


\section*{Appendix}
\label{Section:Appendix}

\section*{Thin Scattering Screen Model}
\label{Section:ScatteringScreenModel}

Pulse broadening is typically quantified by a characteristic timescale, $\tau_{d}$, which is related to the pulsar's distance and scattering measure in the case of scattering from the ISM~\citep{Cordes1991}. We assume that the pulse broadening is produced by multipath scattering through a thin scattering screen, infinitely extended transverse to the line of sight~\citep{Cronyn1970}. Temporal scattering is modeled by a pulse broadening function~(PBF), which describes the electron density in the~ISM. If the electron density fluctuations are characterized by a \text{square-law} structure function~\citep{Lambert1999}, the~PBF is given by a truncated, \text{one-sided} exponential~\citep{Bhat2003}:
\begin{equation}
\text{PBF}(t)=\frac{1}{\tau_{d}}\text{exp}\left(-\frac{t}{\tau_{d}}\right)\Theta(t),
\label{Equation:PBF}
\end{equation}
where $\Theta(t)$ is the unit step function, defined by $\Theta(t$\,$\ge$\,$0)$\,$=$\,$1$ and $\Theta(t$\,$<$\,$0)$\,$=$\,$0$.

We model the unbroadened single pulse emission component as a Gaussian pulse:
\begin{equation}
P(t)=\frac{A}{w\sqrt{2\pi}}\text{exp}\left[-\frac{1}{2}\left(\frac{t-t_{0}}{w}\right)^{2}\right],
\label{Equation:GaussianPulse}
\end{equation}
where $A$ is the amplitude of the pulse, $t_{0}$ is the time of the pulse peak, and $w$ is the intrinsic 1$\sigma$ pulse width. The observed scattered single pulse profile, $P_{\text{obs}}(t)$, is given by the convolution of the intrinsic profile, $P(t)$, with the~PBF in Equation~(\ref{Equation:PBF}) and the impulse response of the instrument, $I(t)$:
\begin{eqnarray}
P_{\text{obs}}(t) & = & P(t)\ast\text{PBF}(t)\ast I(t)\\
& = & P(t)\ast\text{PBF}(t)\ast D(t)\ast S(t)
\label{Equation:ConvolutionResponse}
\end{eqnarray}
The instrumental response, $I(t)$, is derived from the convolution of the impulse response,~$D(t)$, due to incoherently dedispersing the data over a narrow detection bandwidth, and the impulse response, $S(t)$, produced by the radio telescope's detection circuitry~(e.g.,~from a finite sampling time). Here, we ignore the effect of incoherent dedispersion on the observed pulse shape since the \text{intra-channel} dispersion smearing at \mbox{$X$-band} is 25.3\,$\mu$s, which is significantly less than the 512\,$\mu$s sampling time. We also assume that additional instrumental effects are negligible.

The observed pulse shape in Equation~(\ref{Equation:ConvolutionResponse}) has an analytical solution~\citep{McKinnon2014} in the absence of instrumental effects~(i.e., $I(t)=D(t)=S(t)=\delta(t)$, where $\delta(t)$ is the Dirac delta function):
\begin{widetext}
\begin{equation}
P_{\text{obs}}(t)=\frac{A}{2\tau_{d}}\text{exp}\left(\frac{w^{2}}{2\tau_{d}^{2}}\right)\text{exp}\left[-\frac{(t-t_{0})}{\tau_{d}}\right]\times\left\{ 1+\text{erf}\left[\frac{(t-t_{0})-\frac{w^{2}}{\tau_{d}}}{w\sqrt{2}}\right]\right\} +b,
\label{Equation:ObservedPulseShape}
\end{equation}
\end{widetext}
where we have added a constant, $b$, to account for small offsets in the baseline levels of the single pulse profiles. We note that, aside from normalization factors, which can be incorporated into the definition of the amplitude, $A$, our model in Equation~(\ref{Equation:ObservedPulseShape}) differs from Equation~(3) in~\citet{Spitler2014} and Equation~(2) in~\citet{Desvignes2018} by a multiplicative factor of $\text{exp}(t_{0}/\tau_{d})$. This term has a considerable effect on the pulse amplitude when the broadening timescale, $\tau_{d}$, is small.

\bibliographystyle{yahapj}
\bibliography{references}

\clearpage

\end{document}